\documentclass[journal]{IEEEtran}
\usepackage{algorithm}
\usepackage{algpseudocode}
\usepackage[dvips]{color}
\usepackage{times}
\usepackage{subfigure}
\usepackage{verbatim}
\usepackage{amsmath}
\usepackage{amssymb}
\usepackage{amsfonts}
\usepackage{graphicx}
\usepackage{url}
\usepackage{enumitem}
\usepackage{epsfig}
\usepackage{xspace}
\usepackage{thumbpdf}
\usepackage{hyperref}
\usepackage{booktabs}
\usepackage{colortbl}
\usepackage{rotating}
\usepackage{geometry}
\usepackage{pdflscape}
\usepackage{array}
\usepackage{amsmath}
\newcolumntype{L}{>{\centering\arraybackslash}m{3cm}}
\usepackage[noadjust]{cite}

\usepackage{amsmath}
\usepackage{pifont}

\newcommand{\xmark}{\text{\ding{55}}}
\usepackage{verbatim}
\usepackage[edges]{forest}
\usepackage{forest}
\usepackage{lmodern}
\usepackage[lgr,T1]{fontenc}
\usepackage{geometry}
\usepackage[utf8]{inputenc}
\usepackage{longtable}
\usepackage{tikz}
\usepackage{adjustbox}

\begin{document}

\title{Adaptive Bitrate Video Streaming for Wireless nodes: A Survey}

\author{Kamran Nishat,Omprakash Gnawali, Senior Member, IEEE
        and~Ahmed Abdelhadi, Senior Member, IEEE% <-this % stops a space
\thanks{Kamran Nishat, Omprakash Gnawali and Ahmed Abdelhadi are with the University of Houston, Houston, TX 77004, USA
(e-mail:mnishat@uh.edu, gnawali@cs.uh.edu, aabdelhadi@uh.edu)}
\thanks{Manuscript received XXX}}

%\markboth{IEEE COMMUNICATIONS SURVEYS & TUTORIALS,}%
%{Shell \MakeLowercase{\textit{et al.}}: IEEE Journals}

\maketitle

\begin{abstract}

In today's Internet, video is the most dominant application and in addition to this, wireless networks such as WiFi, Cellular, and Bluetooth have become ubiquitous. Hence, most of the Internet traffic is video over wireless nodes. There is a plethora of research to improve video streaming to achieve high Quality of Experience (QoE) over the Internet. Many of them focus on wireless nodes. Recent measurement studies often show QoE of video suffers in many wireless clients over the Internet. Recently, many research papers have presented models and schemes to optimize the Adaptive BitRate (ABR) based video streaming for wireless and mobile users. In this survey, we present a comprehensive overview of recent work in the area of Internet video specially designed for wireless network.  Recent research has suggested that there are some new challenges added by the connectivity of clients through wireless. Also these challenges become more difficult to handle when these nodes are mobile. This survey also discusses new potential areas of future research due to the increasing scarcity of wireless spectrum.
\end{abstract}


\IEEEpeerreviewmaketitle

\section{Introduction}
\label{sec:intro}
\IEEEPARstart{V}{ideo} is the most frequent type of traffic on today’s Internet \cite{1, 2}. It is important for services like Youtube, Netflix, and Facebook to deliver a high Quality of Experience (QoE) during video streaming to sustain revenues \cite{6} and user engagement \cite{7}. Most Internet video delivery services like Twitch, Vimeo,  Youtube use Adaptive BitRate (ABR) to deliver high-quality video across diverse network conditions. Many different types of ABR are implemented in recent years \cite{3,4} to optimize the quality of the video based on different inputs like available bandwidth and delay. Recently, Pensieve, a neural adaptive video streaming platform developed by MIT \cite{3}, it uses deep reinforcement learning (DRL) \cite{8, 9}, and outperforms existing ABRs. 

One of the major challenges will occur in the near future with 5G wide deployment when many devices share the unlicensed spectrum, such as \cite{13, 15, 19} .Video stream applications can be optimized for these resource critical scenarios with the introduction of edge device based feedback to the Reinforcement Learning (RL) based ABR running on the server. 
These edge devices will collect data by spectrum sensing and then allocate the spectrum for the next time slot. This allocation will be transmitted to the ABR server in addition to the feedback from the mobile client.

\begin{table*}
\centering
\caption{LIST OF COMMONLY USED ACRONYMS IN THIS PAPER}
\begin{tabular}{ | c  | c | } 
  \hline
  Acronym & Explanation  \\ 
  \hline
  DASH & Dynamic Adaptive Streaming over HTTP \\
  ABR & Adaptive BitRate  \\ 
  QoE & Quality of Experience    \\ 
  DL & Deep Learning \\
  RL & Reinforcement Learning \\
  5G & $5^{th}$ Generation mobile networks \\
  LTE & Long-Term Evolution \\
  MIMO & Multi-Input Multi-Output \\
  IoT & Internet of Things \\
  HTTP & HyperText Transfer Protocol \\
  DRL & Deep Reinforcement Learning \\
  MAC & Media Access Control \\
  MDP & Markov Decision Process \\
  DQL & Deep Q-Learning \\
  CNN & Convolutional Neural Networks \\
  OFDMA & Orthogonal Frequency-Division Multiple Access \\
  OFDM & Orthogonal Frequency-Division Multiplexing \\
  HAS & HTTP Adaptive Streaming \\
  PoPs & Point-of-Presence \\
  CDN & Content Delivery Network \\
  MPTCP & Multi-Path TCP \\
  EC & Edge Computing \\
  SILP & Stochastic Integer Linear Program \\
  MEC & Mobile Edge Computing \\ 
  MNOs & Mobile Network Operators \\
  KPIs & Key Performance Indicators \\
  DNN & Deep Neural Network \\
  SDN & Software Defined Networks \\
  DRNN & Deep Recurrent Neural Network \\
  \hline
%\caption{ List of abbreviations in alphabetical order.}
\end{tabular}
\label{tbl:acron}
\end{table*}

\subsection{Motivation: Why we need ABR solutions for Wireless Networks?}
Video streaming over wireless/mobile nodes now accounts for more than 70\% of Internet traffic, and it is still growing with a phenomenal rate \cite{1}. Massive deployments of LTE based cellular networks has also played a vital role in this. LTE supports peak down-link bitrate of 300 Mbps, almost 10 times more than over 3G \cite{PiStream15}. However, most of the studies show  QoE is still unsatisfactory \cite{37}.   

New applications of video over mobile client are getting popular \cite{276}. Exponential growth of IoT based networks will increase these innovative scenarios, with the applications like online object detection \cite{566,283} and energy efficient scheduling \cite{291}. Many new applications apply machine learning algorithms like deep learning \cite{519} on video streams on resource constraint mobile devices. These applications introduce new challenges and opportunities for Internet video ecosystem.

\subsection{Prior Survey Articles}
Having established the importance of ABR algorithms optimizing QoE for wireless and mobile clients in particular, in this paper, we are reviewing existing models and algorithms in this area. While,
there exist previous surveys, in the area of Internet video and optimizing applications for wireless networks in our opinion there are none which focuses on mobile video streaming algorithms. Previous surveys like  Seufert et al. \cite{Survey-1} and Bentaleb et al.  \cite{SurveyABRoverHttp}
 discuss different ABR algorithms in general and related influence factors. In another survey by Juluri et al. \cite{Survey-2}, they discussed tools and measurement
methodologies for predicting QoE of online video streaming
services. Similarly in \cite{19AccessSurvey}, authors provide a survey of QoE models for ABR applications. Kua et al. \cite{surveyABR17} focuses on rate adaptation
methods for Internet video in general, provides a comprehensive review of video traffic measurement
methods and a set of characterization studies for well-known
commercial streaming providers like Netflix, YouTube, and
Akamai. The survey in \cite{DLWirelessSurvey} discusses the growing popularity of deep learning (DL) based techniques to solve different wireless network problems. They discuss the applications of DL methods for different layers of the network, but do not include Internet video  and its challenges in particular. Seufert et al. in \cite{surveyVideoQua15} focused on video quality  metrics and measurement approaches that are related to HTTP based adaptive streaming. Similarly, Barakabitze et al. \cite{surveyABR20-2} focused on techniques of maintaining QoE in emerging types of networks based on SDNs and NFVs. They do discuss QoE for multimedia application in LTE and 5G networks but more with the context and opportunities related to SDN/NFV.

Bentaleb et. al. in their survey \cite{surveyABR19} describes a many recent paper related to ABR in detail. Their main focus is a scheme classification based on the unique features of the adaptation logic of ABR algorithms.

Our survey is unique from others in three key aspects: (1) It is focused on clients connected to the internet using wireless technologies like WiFi or cellular network. In all previous surveys none focused on different schemes and their challenges of designing ABR specifically for wireless networks. (2) We provide an in-dept survey of schemes using machine learning in general and RL in particular to optimize QoE of video for wireless nodes. (3) We provide many open challenges in designing ABR for future wireless networks.

\subsection{Contributions of This Survey Article}

In this article, a comprehensive survey on the proposed ABR algorithms for wireless networks is presented.  The  contributions of this review paper are summarized as follows. Towards this end we present in this paper a review of the proposed ABR algorithms for wireless networks. 

\begin{itemize}[leftmargin=*]
	\setlength\itemsep{0em}
	
\item We present and classify the existing works related to ABR for wireless networks. In this paper, we provide an overview of the current state of the art in the field of Internet video in wireless networks.
\begin{comment}
\item We identify some important directions of future research. In this paper, we address challenges in future work to make Internet video more stable on wireless nodes.
\end{comment}
\item We identify some important directions of future research. We present some area where upcoming new standards and their adaptations will create challenges for existing ABR algorithms. We discuss suggestions for future design of ABR algorithms.
\item We review many open-source implementations of different ABR algorithms. And we present their differences and comparisons. 
\end{itemize}

A list of acronyms used throughout the article is presented in
Table \ref{tbl:acron}. The rest of this paper is organized as follows. Section \ref{sec:overview} presents the overview of the Internet video delivery ecosystem. It also introduces basics of machine learning techniques used in the papers discussed in this survey. Section \ref{sec:typeofABR} surveys the bitrate adaptation algorithms for wireless networks. This section is divided in different subsection according to the type of algorithms. Section \ref{sec:openChal} presents different open challenges in the area of ABR for wireless. In the section \ref{sec:opensource} discusses different open-source implementations of ABR algorithms and also different dataset available for experiments. Finally, Section \ref{sec:conc} provides concluding remarks.
 \section{Wireless ABR Video Streaming: An Overview}
 \label{sec:overview}
 \subsection{Why video on wireless is different?}
 
 Internet video systems are designed to cope with the inherent variability in network conditions. Media players  at the client implement ABR algorithms \cite{11, 12, 13}. There are a variety of protocols like MPEG-DASH \cite{DASH11}, Apple HLS \cite{AppleLive}, Microsoft Smooth Streaming \cite{MicroSoftVideo}, Adobe HDS \cite{AdobeLive} etc. that adopt HTTP
based adaptive video streaming. These protocols are called  Dynamic Adaptive Streaming over HTTP (DASH). In these schemes, server splits each video into multiple segments with uniform playback time (typically 1 to 10 seconds). Afterwards, the server encodes these segments into multiple copies with different discrete encoding bitrate levels having 
different sizes. Before a DASH video session starts, a client
obtains the available bitrate map from the server. To download
each segment, the client needs to send an HTTP request
to the server, and specify the bitrate level it prefers for that
segment.

\begin{figure*}[]
	\vspace{-0.2in}
	\centering
	\includegraphics[scale=0.25]{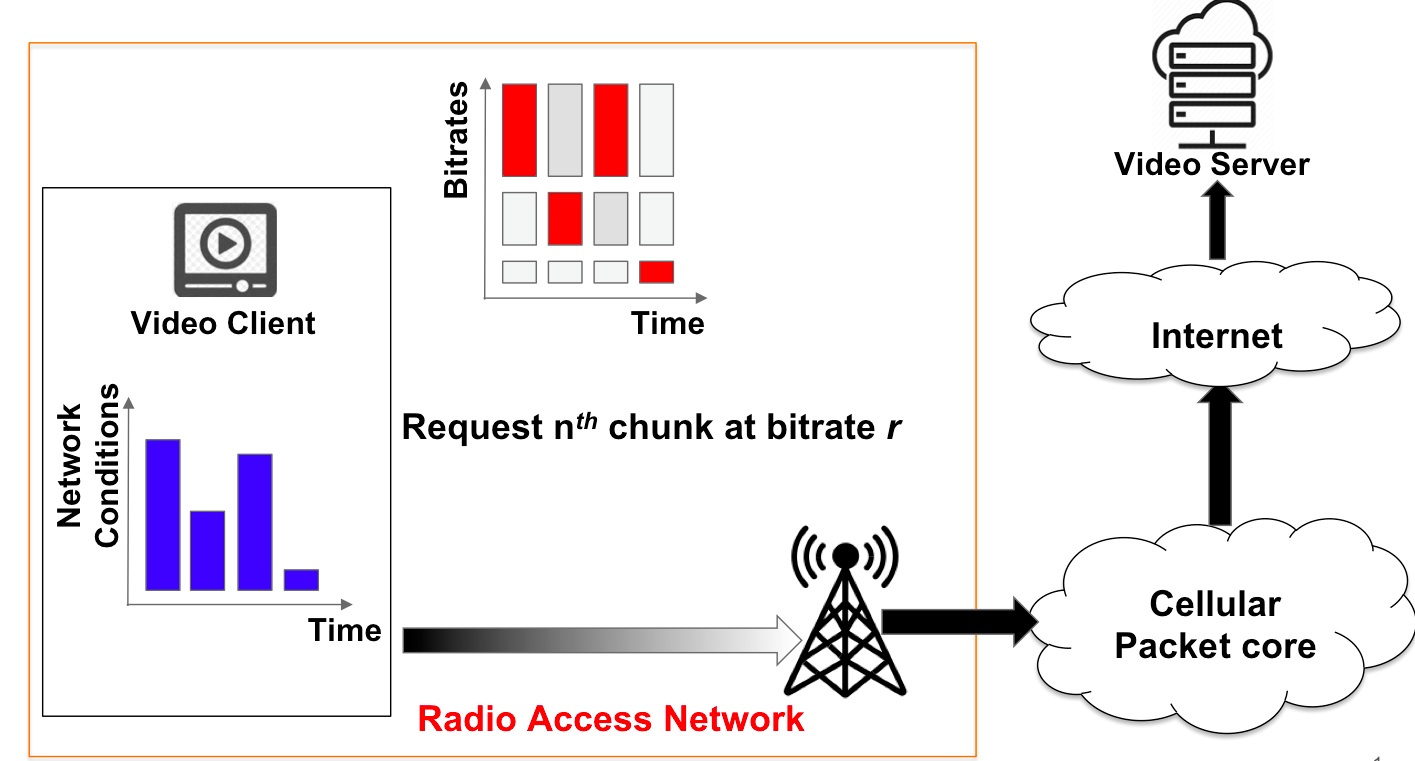}
	\vspace{-0.15in}
		\centering
	\caption{Mobile video player architecture }
	\label{fig:intro}
	%\vspace{-0.1in}
\end{figure*}

Most of the content publishers in today’s Internet serve their videos from some popular content delivery networks (CDNs). These CDNs have point-of-presence (PoPs) in many different geographical and  network domains. By using their PoPs and their peers, CDNs reduce the cost of serving videos and join times, each video is delivered over an ISP network. Most of these Internet service providers (ISPs) have two parts, the core network and the radio-access network (e.g., cellular network) as shown in Figure \ref{fig:intro}. User devices are connected with the ISP via wireless technology like WiFi or LTE. However, these solutions render unsatisfactory performance in WiFi or LTE networks.

ABR algorithms work by (a) chopping the video into chunks, each of which is available at a range of bit rates; then (b) choosing which bit rate to fetch a chunk at based on conditions such as the amount of video the client has buffered and the recent throughput of the network. These ABR algorithms are implemented in video clients. Hence, on a mobile client they must be energy efficient in addition to all other properties like computational efficiency and optimize QoE for the user. 

One of the first ABR algorithms was designed in model predictive control (MPC) \cite{VeyasABR}. It  predicts throughput of future chunk downloads using the historic data of recently downloaded chunks. The predicted value of throughput is used to select the bitrate for the future chunks such that optimizes the QoE function. MPC has a look-ahead window of 5 future chunks. There is an aggressive version of this algorithm called FastMPC which directly uses the throughput estimate obtained using a harmonic mean predictor. 

On the other hand there are algorithms like Buffer  Occupancy  based Lyapunov  Algorithm (BOLA)  \cite{BOLA}  which uses buffer occupancy to  selects bitrate. BOLA solves an
optimization problem to select optimal bitrate. BOLA is a buffer-based algorithm
used in Dash.js \cite{dash}. In contrast to MPC, it does not employ throughput prediction in making decisions. It tries to avoid re-buffering by maintaining a minimum buffer threshold. This threshold can be used to make this algorithm conservative or optimistic about the  future bitrates.

 Recently, RL and other machine learning techniques are used to design  ABR algorithms \cite{DashMMSys16, learningABR2,learningABR3}. Using RL Pensieve \cite{3} was able to outperform the state-of-the-art. Oboe \cite{4} presented an ABR algorithm which performs an automatic tuning of configuration parameter values for each network state independently. This allows Oboe to give better QoE than Pensive.

\subsection{Why Reinforcement Learning is popular for ABR design?}
There are many schemes based on learning-based approach to solve problems in networks in general \cite{18}. Some of them are focused towards applying RL. 
In the past, it has been noted that RL is very suitable to be applied to many computer network problems. RL is quite a natural way to model an optimization control problem \cite{RLSurvey}.     

\begin{figure}[]
%	\vspace{-0.2in}
	\centering
	\includegraphics[width=0.5\textwidth]{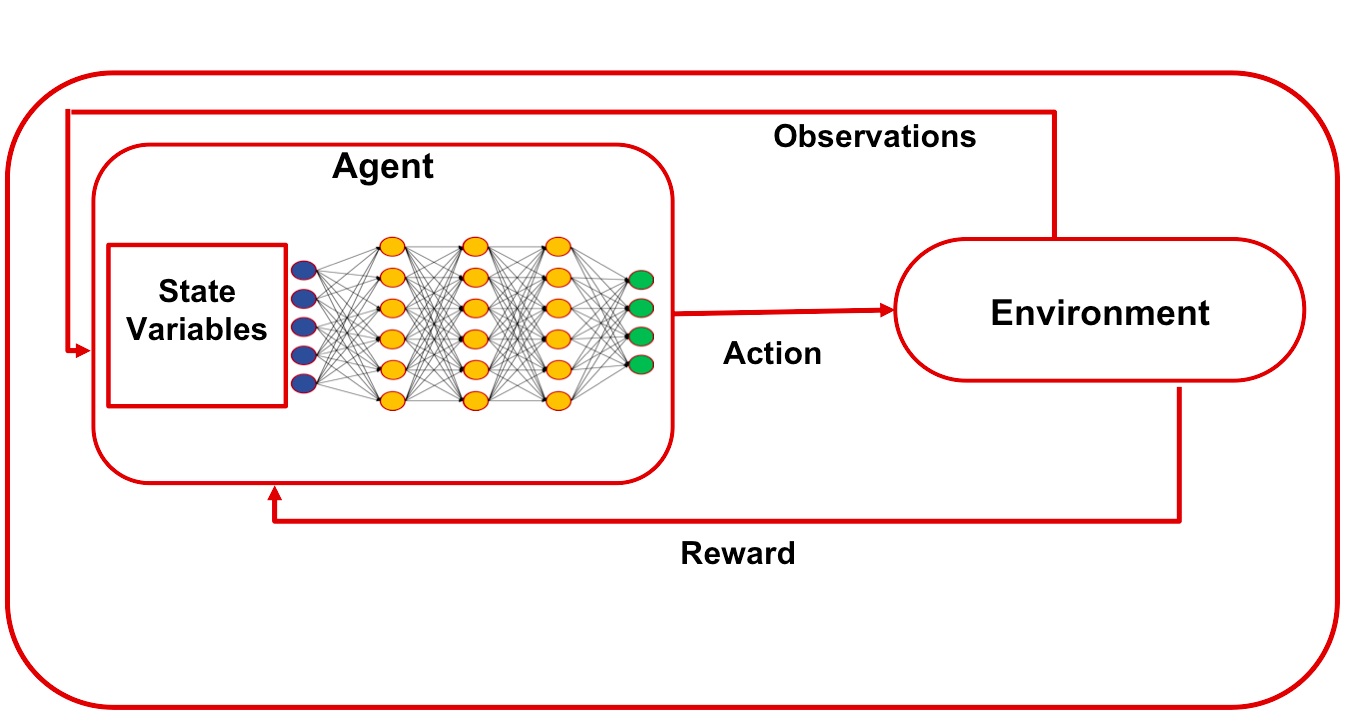}
%\vspace{0.2in}
		\centering
	\caption{Overview of Deep Reinforcement Learning}
	\label{fig:drl}
	%\vspace{-0.1in}
\end{figure}

There are two main entities in a RL problem Figure \ref{fig:drl}, \textbf{Agent} and \textbf{Environment}. Agent observes the \textbf{state} $s_i$ of the environment at each interval and then choose an \textbf{action} $a_i$. Agent receives a \textbf{reward} $r_i$ for his action which can be positive or negative. The goal of an agent in RL is to maximize the cumulative reward defined as follows:

$$ V_{\pi}(s) = E\left[ \sum_{i=0}^{\infty} \gamma^i r_i | s_0 = s \right] $$
 
Where $\pi $ is the \textbf{policy function} $\pi_i(s_i,a_i)$. It gives the probability distribution of the current state and action. While the $\gamma$ is the discount factor for the future reward. Hence, an RL agent learns optimal \textbf{policy} to maximize its rewards. To learn the optimal policy most of the RL applications use Q-learning.
 
In Q-Learning, each pair of state and action $(s,a)$ is mapped to a value under a policy $\pi$. This value is the expected total reward of taking an action $a$ in a state $s$.
$$ Q^{\pi}(s,a) = E\left[ \sum_{i=0}^{\infty} \gamma^i R(s_t,\pi(s_t)) | s_0 = s, a_0=a \right] $$

The goal of Q-learning algorithm is to find the policy to maximize this function. In DRL an agent is learning this optimal policy using a deep neural network (DNN). So, in DRL approximate value functions called deep Q learning function is used. This function is learned by gradient method used in deep learning. Here the agent interacts with the environment like in RL and uses its reward as the training input for the deep neural network. The  goal during training of this DNN is to optimize its  parameters. Hence, it selects actions that can result in the best future return.

Pensive \cite{3} was the first paper to use DRL in designing ABR. Pensive model it using DRL so it can be independent of the assumptions taken by the designer of ABR schemes. In Pensive algorithm, they defined the QoE as the reward of the ABR algorithm working as an agent in the DRL. They modeled their generic QoE based reward function as follows.

\begin{multline*}
QoE={\sum_{i=1}^n{q(R_i)}} - \mu {\sum_{i=1}^i{T_i}}\\
- \sum_{i=1}^{n-1}{|q(R_{i+1})-q(R_i)|}
\end{multline*}
Where the function $q$ is an increasing function of bitrate selected $R_i$ for the interval $i$ so, higher the bitrate higher the reward. The second term depends on the time taken to buffer that segment $T_i$. This will penalize the reward for any re-buffering required before playing the next segment. Last segment of the reward function is penalizing for lack of smoothness. If the bitrate of the video change from the previous segment, then user will observe some lack in smoothness. Pensive \cite{3} is used in many similar research for different variants of the problem. There are two main types of the RL based on the training. First is model based RL and the other is model free RL.

\subsubsection{Model-free Reinforcement Learning}
Model-free RL learns directly from the experiences while in training. The states and transition probabilities of the underling Markov decision process (MDP) are unknown. In ABR, we have no prior model of QoE dependence on the different state variables. Model-free RL learns in more interaction with the environment as compared to the model-based RL Figure \ref{fig:modelRL}.  It is free from the biases of the supervised training data.

\subsubsection{Model-based Reinforcement Learning}
In this type of RL we have a prior model of the system MDP. This will reduce the time of learning and cost in-terms of interactions with the environment, on the other hand, it require designer to provide or learn the model before the start of the training. RL training phase will only optimize or refine this model. 
In this case, if there are inaccuracies in the model, then this will lead to degradation in quality of final output. 

\begin{figure}[]
%	\vspace{-0.2in}
	\centering
	\includegraphics[scale=0.15]{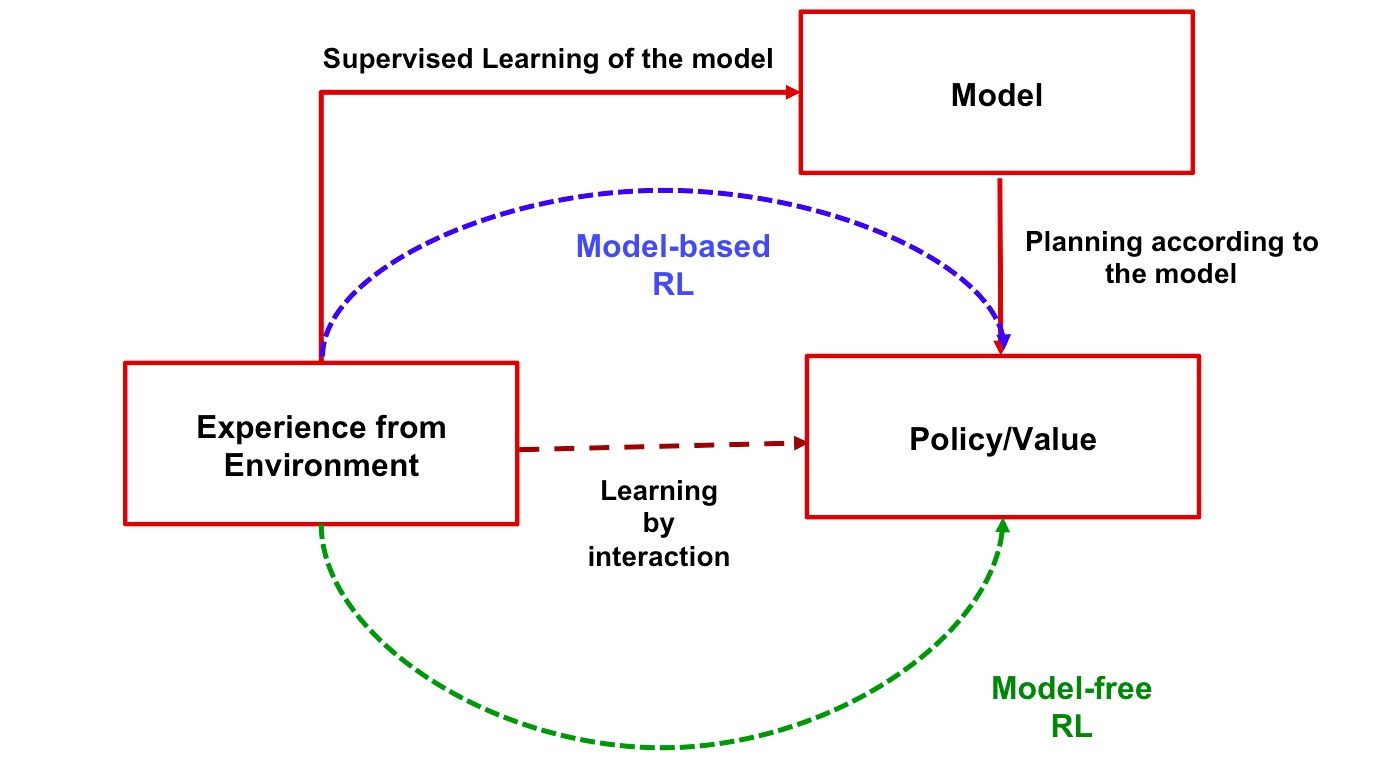}
%	\vspace{0.2in}
		\centering
	\caption{Model-free and Model-based Reinforcement Learning}
	\label{fig:modelRL}
	%\vspace{-0.1in}
\end{figure}

In model-based RL, policy is learned through supervised learning. Then planning over the learned model is done in second phase. Consequently, model-based algorithm uses a reduced number of interactions with the real environment during the learning phase  \ref{fig:modelRL}. So, learning can be much faster because there is no need to get the feedback from the environment.  On the downside, however, if the model is inaccurate, we risk learning something completely different from the reality.

\section{Different types of ABR algorithms}
\label{sec:typeofABR}
There are different types of research in this area. Some have designed ABR algorithms for mobile nodes to incorporate the movement of the nodes, while others focused on the resource constraints like spectrum and energy of the video client. 

Mobile Network Operators (MNOs) offer different packages to increase their number of users. In some packages, they offer not to count certain services like Facebook, WhatsApp and Netflix toward monthly data quota. But they limit the rate of the users toward those services. Traditional ABR based services does not account for this rate limiting. Here traditional throughput maximization based ABRs will not perform well. Zero-rated QoE \cite{Zero-ratedQoE} proposed a novel approach which uses the collaboration of the content provider and MNOs. They designed an ABR which focuses on improving QoE in these special scenarios. They implemented their approach in a simulated environment and performed evaluation with the baseline.

\subsection{Machine Learning based ABR schemes}
Comyco \cite{10} is another study using the Learning-based ABR. They discuss a few weaknesses of previous RL-based ABR algorithms. Their measurement study shows that the quality of video presentations is not always maximized by QoE metrics based on only video bitrates, re-buffering times and video smoothness. They proposed \emph{Imitation Learning} instead of supervised learning can address these weaknesses. 

Reinforcement learning is also applied to make video streaming appear more smooth to the user.% Please add the following required packages to your document preamble:
% \usepackage[\xmarkrmalem]{ulem}
\newgeometry{margin=1cm} % modify this if you need even more space
%\begin{landscape}
% \usepackage[\xmarkrmalem]{ulem}
\begin{table}[p]
\vspace{-2.5in}
\centering
\hspace{2.1in}
 \caption{A SUMMARY OF RELATED PAPERS AND MAIN IDEA.}
    \label{tab:summary}
\begin{tabular}{|l|l|l|l|}
\hline
{  \textbf{Paper}} & {  \textbf{Main Idea}} & {  \textbf{Is ABR?}} & {  \textbf{For mobile?}} \\ \hline
{  FESTIVE   \cite{VeyasABR}}             & {  Stateful ABR with randomized chunk   scheduling to avoid synchronization biases}    & {  \xmark}               & {  \checkmark }                       \\ \hline
{  Pensive    \cite{3}}                    & {  DRL is used to optimize the QoE}                                    & {  \xmark}               & {  \checkmark }                       \\ \hline
{  Oboe    \cite{4}}                       & {  Combine offline and online tunning of the parameters}                               & {  \xmark}               & {  \checkmark }                       \\ \hline
{  pstream \cite{PiStream15}}                 & {  Improves the QoE by taking   advantage of the PHY information of LTE networks}      & {  \checkmark }              & {  \checkmark }                       \\ \hline
{  MP-DASH \cite{MP-DASH}}                 & {  Use MPTCP to schedule}                                                              & {  \checkmark }              & {  \checkmark }                       \\ \hline
{  Wi-Fi Goes to Town  \cite{42}}          & {  Improves the QoE during high speed handovers}                                       & {  \checkmark }              & {  \xmark}                        \\ \hline
{  HotDASH \cite{HotDASH}}                 & {  Use DRL to detect user specific important part of video to improve their   quality} & {  \xmark}               & {  \checkmark }                       \\ \hline
{  Zero-rated QoE  \cite{Zero-ratedQoE}} & {  Collaboration of MNOs and content providers to improve QoE for rate   limited users} & {  \checkmark }              & {  \checkmark }                       \\ \hline
{  QARC    \cite{QARC}}                    & {  Imrove the perceptual quality of the video instead of traditional QoE metrics}      & {  \xmark}               & {  \checkmark }                       \\ \hline
{  Bursttracker \cite{37}}                 & {  Find the bottleneck in the video streaming over the LTE network}                    & {  \checkmark }              & {  \xmark}                        \\ \hline
{  Qflow \cite{qflow}}                     & {  Used both Model based and Model free DRL}                                           & {  \checkmark }              & {  \checkmark }                       \\ \hline
{  NAS \cite{NAS}}                         & {  A deep neural network based ABR}                                                    & {\checkmark   }                 & {  \checkmark }                       \\ \hline
{{  IncorpPred   \cite{IncorpPred}}} & {  Incorporate cellular throughput prediction to improve ABR}                                  & {\checkmark   }                 & { \checkmark  }                          \\ \hline
{  Comyco \cite{10}}                   & {  Proposed imitation based Learning instead of supervised learning}                   & {  \xmark}               & {  \checkmark }                       \\ \hline
{  Jigsaw \cite{Jigsaw}}                   & {  4K video streaming}                                                                 & {  \checkmark }              & {  \xmark}                        \\ \hline
{  TransPi \cite{TMC19Hardware-assisted}}  & {  Introduced hardware-assisted video transcoding for Wireless}                        & {  \checkmark }              & {  \checkmark }                       \\ \hline
{  CASTLE \cite{CASTLEAir19}}              & {  Client schedular to minimizes Load and Energy at the same time}                     & {  \checkmark }              & {  \checkmark }                       \\ \hline
{  ACAA \cite{19ContentAwareQoE}}           & {  Incorporate  user’s subjective   viewing information to improve ABR}                & {  \xmark}               & {  \checkmark }                       \\ \hline
{  LinkForecast \cite{LinkForecast}}       & {  Bandwidth prediction for LTE network}                                               & {  \xmark}               & {  \checkmark }                       \\ \hline
{  QUAD \cite{QUAD}}                       & {  Reduce the bandwidth usage while maintaining high QoE}                              & {  \checkmark }              & {  \checkmark }                       \\ \hline
\end{tabular}
\end{table}
%\end{landscape}
\restoregeometry  In \cite{PowerRLforVBR}, the authors design an optimization problem for ABR to exploit power control over multiple sub channels at the transmitter in such a way that video quality remains smooth.It penalizes both for buffer underflow and overflow.  Then, they mapped this constraint optimization problem into a MDP. The MDP is solved using reinforcement learning techniques.

It is challenging to implement heavyweight ARB techniques in resource constraint mobile devices. PiTree \cite{PiTree} introduced the idea of using lightweight decision trees to simplify the complex and heavyweight neural network based techniques. PiTree give a highly scalable framework to convert complex ABRs into decision trees. They also provide some theoretical upper bound on the optimization loss
during the conversion.

Challenges of Video streams with high quality are increased in the case of remote drone piloting. The study in \cite{16} discusses these challenges. It suggests decreasing the coupling of different functional blocks. They proposed to use edge-computing elements in addition to adapting for network conditions.

QFlow \cite{17} paper used reinforcement learning to perform one-way adaptive flow prioritization at the edge network. QFlow argued current link are application agnostic in their scheduling. By making these links intelligently adapt for different type of traffic leads to a better QoE of the video streaming. QFlow borrowed concepts of network level priority queues from software defined networks (SDNs) and apply it to PHY/MAC layer using Software-Defined Radios.

QFlow uses RL to optimize the QoE for video by adapting configurations. QFlow  uses both model-free and model-based RL approaches. 

According to their evaluation, RL based approach not only improves the QoE but also results in better buffer state and lower stall duration.
\begin{comment}

It is challenging to implement heavyweight ARB techniques in resource constraint mobile devices. PiTree \cite{PiTree} introduced the idea of using lightweight decision trees to simplify the complex and heavyweight neural network based techniques. They give a highly scalable framework to convert complex ABRs into decision trees. They also provide some theoretical upper bound on the optimization loss during the conversion.
\end{comment}
The survey \cite{18} provides a comprehensive survey of deep learning-based techniques used in different wireless networking scenarios. It also highlights some potential applications of DL to networking, like in network security and  user localization.

In QARC \cite{QARC} they have designed a rate control algorithm that is focused on the perpetual quality of the video. The perpetual quality, of the video is defined as how many objects are in the image and how bright or dark it is. For low perceptual quality parts of the video we can save the bandwidth and delay by requesting video at the low quality. While for high perceptual quality, parts should be downloaded at a higher bitrate. This can be achieved by lower sending rate and latency. QARC also use DRL to train the neural network to predict the future video frames based on the perceptual quality of the previous frames. It employs two-fold training of DRL one for prediction of perceptual quality and the second one using A3C based asynchronous training technique to train the actual RL  algorithm. They did trace driven analysis of their techniques and compare it with Google Hangout and compound TCP.

\subsection{Egde computing based ABR systems}
\begin{figure*}
%	\vspace{-0.2in}
	\centering
	\includegraphics[width=\textwidth,height=6cm]{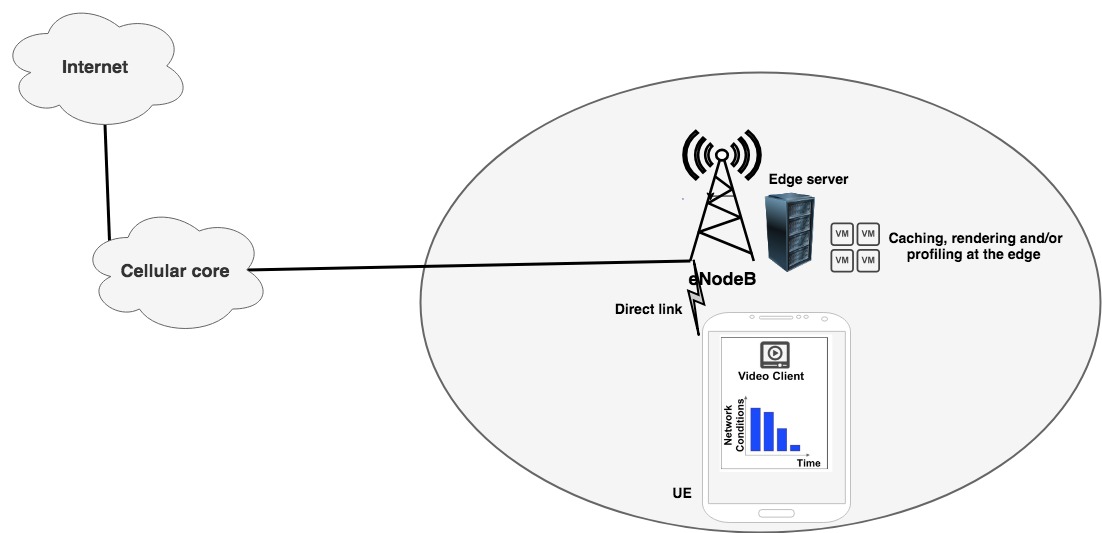}
	\vspace{0.2in}
		\centering
	\caption{Introduction of MEC to improve the ABR based video streaming}
	\label{fig:Edge}
	%\vspace{-0.1in}
\end{figure*}

Increasing demands of lower network delay and  higher data transmission rate are getting difficult  meet from traditional ABR systems. Recently, edge computing (EC) Figure \ref{fig:Edge} based optimizations to ABR systems have been proposed to meet these challenges. In the paper \cite{EdgeBilal}, authors present some of the challenges and limitations of the current ABR applications. They proposed an edge computing based solution to address these challenges and limitations. QFlow \cite{17} is an Edge computing based ABR system using ML based model to optimize QoE.

ShareAR \cite{EdgeShareARAR19} is a multi-user  augmented reality (AR) system which uses edge nodes to optimize QoE for the user. The main challenge in multi-user  AR is the communication between AR platforms. There are no prior work involving data transmission in between AR devices and their impact of the QoE. In multi-user AR, devices can have different fields-of-view. They need to render their respective FoVs. In their system, they overcome these challenges and  implemented a prototype of the system using two Android devices and an edge server.

In FlexStream \cite{EdgeFlexStream} they leveraged the SDN functionality to get the benefits of centralized management of distributed components. Here they use wireless edge device like AP as SDN controller. They have implemented their system as a light weight controller. In there evaluation, they showed FlexStream can achieve appropriate bandwidth
distribution.

Blockchain technology is used by decentralized peer-to-peer video streaming systems to monetize using smart contracts. In these new video streaming systems, content creators, consumers and advertisers can communicate with each other without the help of a trusted third party. There some challenges in designing these systems like processing and publishing of the video content in these systems. In \cite{EdgeBlockChain}, they propose using edge computing servers to offload these computationally intensive tasks. They proposed to employ edge servers through distributed block-chain based incentive mechanisms.

Mobile Edge Computing (MEC) is getting popular to provide low-latency ABR service. One way to decrease the latency of the system is to use MEC servers for video caching. Tran et al. \cite{EdgeVideoCaching} investigates a novel caching scheme using multi-server MEC systems. Their systems use two timescales. They formulated stochastic integer linear program (SILP)  to integrate these two timescales of long-term caching and short-term video retrieval mode. By using simulations they showed the effectiveness of their system by reducing access delay and increasing cache hit ratio.

PrivacyGuard \cite{EdgeExtremeSDN} is a system designed and developed to obfuscate the activities of sensitive
IoT and mobile applications from attacks over WiFi network. They have implemented a prototype this systems on Android mobile devices that to apply application level traffic shaping and IP-sec tunneling schemes.

\subsection{ABR with different optimization goals}

In Wi-Fi Goes to Town \cite{42}, implement a WiFi based hotspot network using picocell size access point networks along the road to support vehicular communication over high speed. They implemented optimized version of IEEE 802.11k and 802.11r standards. Although their main focus is not video streaming but most of their evaluation is done over video. Their scheme provides more reliable video stream for high speed mobile client. Also it improves the QoE metrics for the video like rebuffed ratio. 

Sengupta et al. in HotDASH \cite{HotDASH} focused on improving the video quality for the specific user requirements. In most of the video streaming situations there is some content of the video which is more important for the user. They use DRL to detect that part of the video and then the requirement, therefore, is for a video streaming strategy to into account the content preferences of the users. So, ABR will be aware of the high-priority temporal content. ABR try to pre-fetch those high priority parts of the video at much higher bitrate. HotDASH maximizes the content preferences of users, in addition to optimal use of bandwidth. They implemented their scheme in dash.js and compared it with the six baseline algorithms like FESTIVE, FastMPC and PENSIVE.

Most of the ABR algorithms are not designed with the consideration of data consumption. But most cellular customers have limited data in their monthly data plan. According to  \cite{DataPlanStat} average U.S. cellular customer has only 2.5 GB per month data plan, while one hour high definition (HD) video on mobile require 3 GB data. QUAD \cite{QUAD} focuses on reduce the bandwidth usage while maintaining high QoE for the user. Their scheme is also energy efficient because it requires to download less amount of data. 

QUAD introduced a novel Chunk Based Filtering (CBF) approach which leverages two fundamental tradeoffs of video quality and bitrates. First, higher bitrate leads to diminishing return in terms of video quality. Second, different chunks have different impact on video quality. Their scheme selects chucks to maximize the QoE while keeping the data consumption minimal. QUAD implemented its scheme in both dash.js and ExoPlayer and perform evaluations. They compared their approach with RobustMPC and PANDA.

At the same time MP-DASH \cite{MP-DASH} take a different approach to optimize video quality over mobile devices. Their focus is on leveraging the availability of multi-path in many common mobile devices like cell phones. They have WiFi card and LTE modem at the same time. So, in many cases it is possible to use LTE opportunistically. They used Multi-Path TCP  (MPTCP) to implement their approach. It prefers WiFi over LTE when at home. The evaluations were done in both controlled setting and in the wild. Trace of throughput and RTT of the WiFi networks at 33 locations in the US for the evaluation. MP-DASH is impelled using GPAC. FESTIVE and BBA is implemented over the multi-path scheduler. In \cite{QualityAwareTrafficOffloading} they implemented traffic offloading build on the MP-DASH approach  for general application. 

ACAA \cite{19ContentAwareQoE} is a scheme focused towards semantic information of video content. Recently ABR researchers are  designing with user’s subjective viewing information to improve the QoE of the video specific to the user requirement. ACAA use the research on video affective content analysis. It incorporated individual user preferences into the bit-rate adaptation decisions to improve the QoE. Identify the user relevant parts of the video and then assign bit-rate budget according to it. They compared their scheme with BBA and buffer-based adaptation (BBA), and model predictive control (MPC). ACAA implemented their scheme with the DASH client in accordance with \cite{traceDASH} to perform trace-driven evaluation platform with python 2.7.

\subsection{Measurement of different ABR schemes}
Many papers study performance of different ABR algorithms and make a comparative study. Some of them performed active measurement \cite{active1} \cite{ConfusedIMC12} while other perform passive measurements \cite{passive1}.  There are others like \cite{21} and Puffer \cite{22}  made a database of different ABRs.
Duane et al. developed the Waterloo SQoE-III database \cite{21}. This database provides a subjective evaluation of different QoE models and ABR algorithms. SQoE-III evaluated Rate-based, AIMD, Dynamic Adaptive Streaming algorithm, etc in their paper. According to their evaluation 5 out of 6 models are quite close in terms of performance. In addition to the experimental evaluation of other ABR techniques, Puffer \cite{22} developed a live TV streaming website. This prototype website has attracted over 100,000 users across the Internet. This system works as a randomized experiment; one set of ABR schemes is randomly assigned to each session.

Haung et al. \cite{ConfusedIMC12} is the first study performed in this area. In their study, they perform a measurement study of three popular video services Hulu, Netflix, and Vudu. 
\cite{ABRoverQUICMeas} studied and discussed the impact of QUIC on QoE of the popular ABR. It also discussed how can existing ABRs leverage the potential benefits of QUIC.

The study in \cite{MobileMeasre1} is a general measurement study of application performance in the rapid deployments of LTE networks. Data traces has been used from different major LTE providers. VideoNOC \cite{active1} is an passive measurement study of Video QoE for Mobile Network Operators (MNOs). VideoNOC presented an approach to assess the QoE for different MNOs using objective metrics for video quality. To get an objective estimate of QoE metric they collected HTTP/S traffic in the core of the LTE network. VideoNOC performed many efficient and scalable cross-layer analytics over these logs.

Recently, a third-party based system to is designed to evaluate and understand the behavior of different closed source ABR based streaming services \cite{CSIMeasureABR}. Channel State Information (CSI) can also able to understand the behavior of ABRs in the presence of traffic encryption. 

\subsection{4K and 360-degree video streaming}

4K videos are now getting increasingly common. New applications like virtual reality (VR) and augmented reality (AR) will make 4K extremely important in the coming future. These applications do not only require high resolution but also very low latency. In there raw form 4K video stream requires more than 2Gpbs physical data rate. Currently, IEEE 802.11ad based WiGig card are commodity wireless cards supporting these data rates. These devices work in 60GHz spectrum. In this range, transmission is highly sensitive to mobility. There can be drastic change in the throughput for minor movement. In the case of blocking, the line of sight throughput might be affected and reaches to zero.

In the presence of these large throughput variations, traditional video codecs like H.264 and HEVC become infeasible. To overcome these limitations, layered video codecs are used by Jigsaw \cite{Jigsaw}. It uses scalable video coding (SVC) which is an extension of the H.264 standard. The study in \cite{Jigsaw} use    fast encoding schemes and implement it using  new layered video coding methods.

Panoramic video is another emerging application of video streaming, it is known as  $360^{\circ}$ video. Platforms like Facebook and YouTube also support them.
Flare \cite{Flare360Video}, presented a practical prototype of a $360^{\circ}$ video streaming solution using commodity devices. This study predicts future behavior of the user to fetch only the relevant portions of the video to cover the view of the user, which enables Flare to reduce the bandwidth usage of the system significantly. 

This viewport-adaptive $360^{\circ}$ streaming is an establish technique. Flare is the first complete working implementation on commodity mobile phone. It uses a online machine learning (ML) algorithm to predict head movement of the user which changes the users’ future viewpoint.

Another $360^{\circ}$ video streaming system is presented in Rubiks \cite{Rubiks360Video}. Rubiks discusses different challenges of implementing tile-based video streaming techniques used in different implementations to predict field of view (FoV) of the user.  In resource constraints of commodity smartphones, it is not possible to meet with the requirements these tile-based systems. 

Rubiks uses HEVC to implement tile-based streaming instead of H.264 \cite{h264} used by previous system. HEVC \cite{hevc} has a built-in tiling scheme to encode video data. Their system can stream different parts of the video at different bitrates. This allows them to download tiles in different quality according to their probability of viewing. Managing the amount of data downloaded at the client it can control the decoding time. Decoding time increases substantially for 8K videos on mobile device. 

In the paper \cite{DRL360}, DRL is used to implement a panoramic video streaming system. Their system used DRL to optimize QoE using a broad set of features. Here their focus is on two main challenges of 360-degree video. First, there is a large number of time-variant features which needed to be adapted to achieve a reasonable quality. Second, QoE metrics are also different for different scenarios. 
Zhang et al. uses DRL to find an optimization model. This model finds the best rate allocation scheme for different scenarios.

Tang  et al. in  \cite{OnlineBitrate360} presented a promising approach to improve QoE  for the user in a 360-degree video streaming system. Their focus is on a streaming a newly generated 360-degree video. In this case, there is no historical viewing information available which can be used to predict user viewing behavior. 
In these scenarios, there are additional challenges of learning FoV patterns online and also the lengths of these FoV segments are also unknown in advance. The authors present OBS360 algorithm which is an online bitrate selection algorithm to optimize the user’s QoE in 360-degree video.  OBS360 algorithm is able to learn user’s FoV preference and also the time-varying downloading capacity of the user.

Perfecto et al. \cite{360MultiUser} they discussed  Immersive virtual reality (VR) applications. These applications require achieving motion-to-photon (MTP) delays, which are defined as end-to-end latency of 15-20 milliseconds. Providing $360^{\circ}$  video with these delay guarantees is quite a challenge. They applied a deep recurrent neural network (DRNN) to predict the upcoming tiled FoV. In addition to that, they exploit millimeter wave (mmWave) multicast transmission at the physical layer to improve the efficiency of the system.

\subsection{Video stream in vehicular networks}
Video streaming in vehicular networks is even more challenging \cite{25}. There are different types of communications in V2X networks. In recent years, many papers try to address these challenges \cite{26,27}. Some of them use Edge network-based caching techniques and other explored learning-based approaches to optimize video streaming. 
 
Recent papers explore the use of different machine learning-based approaches to optimize video streaming in wireless networks \cite{29,30, 31, 32}. Some of them focus on the presence of IoT devices on the same spectrum, others optimize for energy-efficiency. In \cite{28}, they perform an experimental analysis of 10 widely deployed ABRs. Their measurement shows none of the deployed ABRs focus on available bandwidth and some leave a large fraction of available network capacity unused.

\subsection{Optimizing video during handovers}
Wi-Fi Goes to Town \cite{42}  was one of the first research approaches to implement a performance-tuned version of the IEEE 802.11r and 802.11k fast handover protocol. It try to use Picocells to increase the capacity of the network, which results in higher spectral efficiency and throughput. In \cite{42} focus is not on video delivery. Focus of \cite{42} is on maximizing throughput that can often lead to lower QoE for video clients.  Wang et al. \cite{41} propose a real-time handover protocol called mmHandover for a 5G network working in mmWave. 
In the past, there have been many efforts to design mechanisms to predict handovers \cite{45, 43 ,44}. These schemes try to predict handovers to improve the QoS of different traffic on mobile devices. 

\subsection{Understand the network bottleneck to improve ABR}

In their paper BurstTracker \cite{37} focused on LTE networks. BurstTracker identify issues affecting the QoE of the video streaming performance. BurstTracker is able to identify a surprisingly different bottleneck. 
BurstTracker understand the scheduling pattern of the LTE base station. Their focus is to identify the occupancy of each user’s download queue. If this queue is not empty for one scheduling cycle then access link the bottleneck link. It means that data was in the queue of LTE base station and was not being delivered in the next cycle. If there are so many cycles like that, they will decrease the QoE for the video. 

\begin{figure}[]
%	\vspace{-0.2in}
	\centering
	\includegraphics[width=\columnwidth]{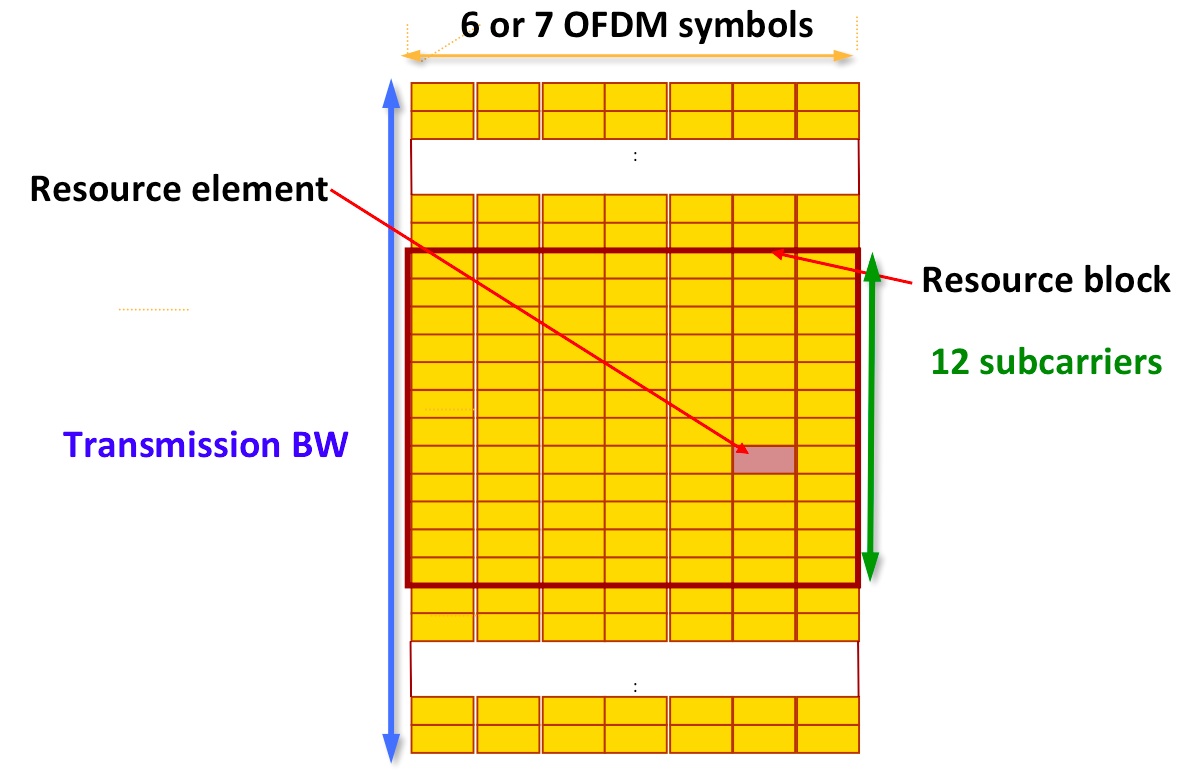}
	
	%	\centering
	\caption{Resource Block and Resource Element in LTE networks}
	\label{fig:RB}
	%\vspace{-0.1in}
\end{figure}

One of the main challenges, is that user queue information is not available at the client. So, this approach designed a method to estimate this information. The approach observes that if a user is selected for transmission and its queue is full, then the LTE base station scheduler allocates the complete millisecond duration resource block (RB). As shown in the Figure \ref{fig:RB} a RB is the smallest unit of resources that can be allocated to a user. It consists of 180 kHz in frequency while 6-7 OFDM symbols in time. In frequency it is further divided into 12 sub-carriers of 15 kHz each. Using these insights BurstTracker is able to find out most of times user queue becomes empty before the complete allocation of the user queue. This suggests that bottleneck is not at the base station radio link. According to BurstTracker, most of the large LTE network providers use split-TCP middle boxes. Due to TCP slow start used by middle boxes for TCP connections, these middle boxes introduce serious performance bottleneck.  

 PiStream \cite{PiStream15} has determined total download resources allocated to the user on a LTE base station. BurstTracker is able to estimate it at the user and then use it to improve the estimation of bitrate. Pistream assumes in case of the bottleneck it is at the base station.

\subsection{ABR algorithms with cross-layer optimizations}
In the ground breaking paper \cite{PiStream15} PiStream the first presented a challenge faced by DASH players in LTE. LTE bandwidth is very high, around 10x than its predecessor 3G. Despite this high bandwidth video clients does not perform well. PiStream motivated this problem with a measurement study which shows how a DASH client behaves in the LTE network. DASH uses a throughput estimator to predict the data rate. But in the LTE network this estimate is mostly an underestimate. This leads DASH selecting a lower quality for the future. PiStream observe that in the LTE networks all the bandwidth information for the access link is known to the LTE network. The approach takes advantage of the Physical layer information to get an accurate estimate of the bandwidth. 
PiStream implemented their scheme using SDRs at the Physical layer and using GPAC player as the open source client. In their evaluations, they compared their scheme with FESTIVE,  BBA, and  PANDA.

Recently,  Raca et al. \cite{MMsys19LTEMesurement} designed a approach to address the challenges to ABR video in the challenging environment of cellular network. They observed in cellular network radio channel, conditions and load on the cell are continuously changing. In addition to that, there is gap in the time scale among different components of the system. Transport layer protocols react at the granularity of hundreds of milliseconds, while radio channel changes at the fraction of millisecond. One the other side, base station can allocate resource in a bursty manner. ML based approach is used to predict throughput of the mobile devices in a LTE network. There ML model is learned on cellular trace data. Their approach shows the importance of radio level metrics in the video streaming applications \cite{MLTCPthr}. Their technique is implemented in Android video player (ExoPlayer) and performed evaluation on a real testbed. 

In \cite{MMSys20}, a dataset for 5G measurements is presented. These measurements are performed on a a major Irish mobile operator. In this dataset all the key performance indicators (KPIs) for client-side cellular metrics and throughput are collected. Dataset is collected with two different mobility patterns of the user driving and static.  Also, stream content is generated from Amazon Prime and Netflix streaming device. In this dataset, GPS based location information is also present. This data is generated from Android  network monitoring application, G-NetTrack Pro. This application can run over a non-rooted Android phone.

In addition to the real dataset, a second synthetic dataset is also presented in \cite{MMSys20} . This data set is generated from  ns-3 \cite{ns3} using  a large-scale multi-cell 5G/mmwave framework.

One of the first cross-layer ABR algorithm for wireless network was CrystalBall algorithm \cite{VideoCrystalBall}. It is a two step algorithm to predict available bandwidth. Main ABR algorithm is based on it. In this study, the effects of the prediction quality on the accuracy of the ABR. CrystalBall shows with error mitigation techniques some level of prediction error can be tolerated. 

Another similar technique is CQIC \cite{HotMobileCrossLayer} to predict TCP throughput using Radio level information in smart phones. 

Yue et al. \cite{LinkForecast} they showed even a trace of 500 data points can be used to develop an accurate model of LTE link level predictions using a ML model. LinkForecast presented an extensive measurement study to understand the bandwidth allocation algorithm of current cellular networks. Most of them allocate a fair proportion of the available bandwidth. Available bandwidth is calculated using the observations of the recent past throughput and link conditions. Using this measurement study as motivation which shows benefits of sharing application layer throughput to the lower-layer. LinkForecast explore the idea of sharing lower-layer link information to the application. 

LinkForecast designs a ML based framework to predict link bandwidth in real time. This framework combine both upper and lower layers information for future prediction. According to their evaluation this technique is not only more precise but also lightweight and insensitive to the training data. 

In \cite{MMsys19LTEMesurement} authors investigate further on the observation of high variability of network conditions in  cellular radio access networks. Publicly available data sets \cite{Dataset2} are used to learn a machine learning model. The data set is very rich in terms of parameters and different mobility patterns like static, walking, car and train etc. Random forest based technique is selected as their ML model. They used random forest (RF) as model for prediction. By increasing the number of trees and using mean of all trees in the RF as the predicted value avoid over-fitting. 

Parameters used in \cite{MMsys19LTEMesurement} model are available through Android Debug Bridge (ADB) APIs. Implementation is done on ExoPlayer and performed evaluation on a real testbed. In there evaluation, they show the effectiveness of their system by improving all QoE metrics.

%\newgeometry{margin=2cm} % modify this if you need even more space

%
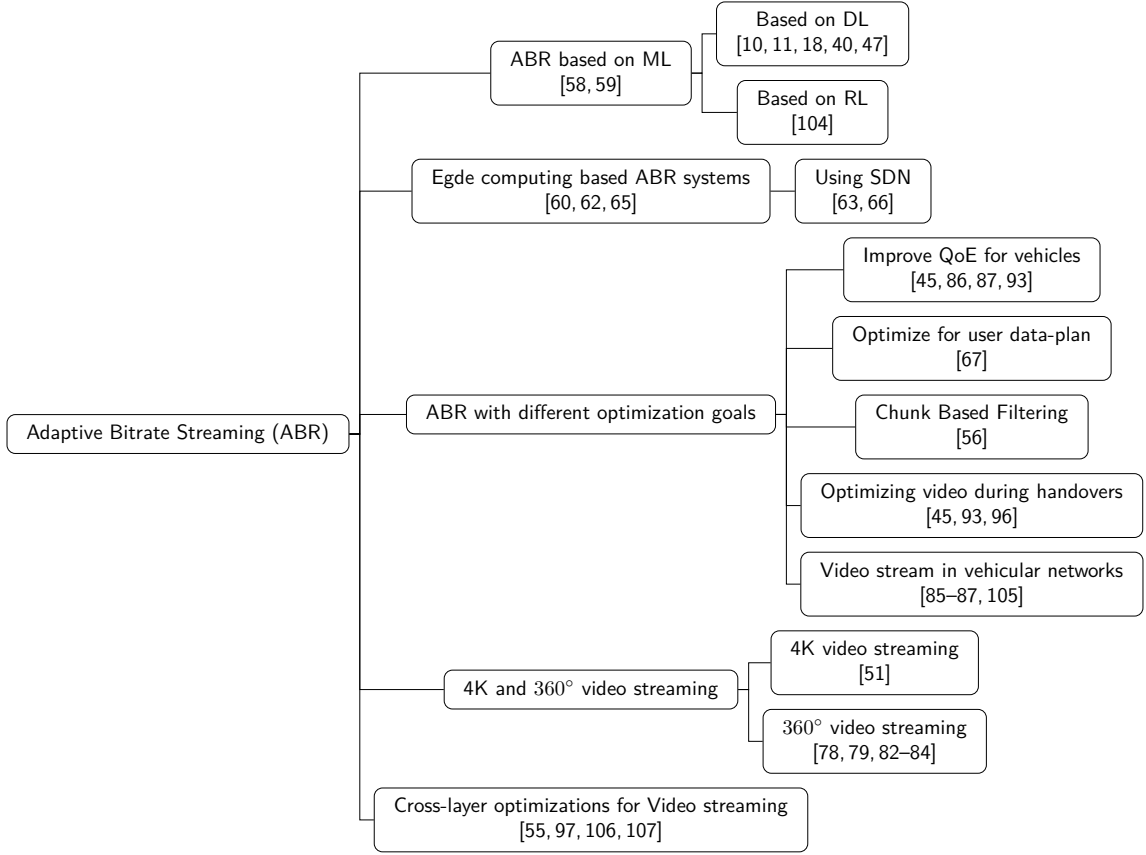
\begin{figure*}
\begin{adjustbox}{width=\textwidth}
\centering

\begin{forest}
  for tree={
    font=\footnotesize,
    grow'=0,
    draw,
    align=c,
    font=\sffamily,
    rounded corners,
    parent anchor=east,
    child anchor=west,
    edge path={%
      \noexpand\path [\forestoption{edge}] (!u.parent anchor) -- ++(5pt,0) |- (.child anchor)\forestoption{edge label};
    }
  },
  [{Adaptive Bitrate Streaming (ABR)\\}
      [{ABR based on ML\\ \cite{PiTree,16} }
      [{Based on DL \\ \cite{15,19,519,18, QARC}}]
      [{Based on RL \\ \cite{PensieveSIGCOMM}}]
     ]
    [{Egde computing based ABR systems \\\cite{EdgeShareARAR19,17,EdgeVideoCaching}}
      [{Using SDN \\ \cite{EdgeExtremeSDN,EdgeFlexStream}}]]
      [{ABR with different optimization goals\\}
        [{Improve QoE for vehicles \\  \cite{42,26,27,41}}]
        [{Optimize for user data-plan \\ \cite{DataPlanStat}}]
        [{Chunk Based Filtering \\ \cite{QUAD} }]
        [{Optimizing video during handovers \\ \cite{41, 42,44}}]
        [{Video stream in vehicular networks \\ \cite{24,25,26,27}}]
       ]
      [{4K and $360^{\circ}$ video streaming}
        [{4K video streaming \\ \cite{Jigsaw}}]
        [{$360^{\circ}$ video streaming \\ \cite{Flare360Video,Rubiks360Video,DRL360, OnlineBitrate360,360MultiUser}}]
      ]
      [{Cross-layer optimizations for Video streaming \\ \cite{LinkForecast,MMsys19LTEMesurement,MagzineThroughputPrediction,EnergyAwareABR}}]
    ]
\end{forest}
\end{adjustbox}
\caption{Major categories of ABR for wireless networks in the literature}
\label{typesABR}
\end{figure*}

Chen et al. \cite{EnergyAwareABR} authors proposed ABR can save energy during video stream if it consider the context of streaming. It means if user is watching a video in a room may have completely different QoE requirements as compared to a user on a moving vehicle. Using traces they have modeled the impact of vibration level in addition to video bitrate on the QoE and signal strength on power consumption. They designed an optimal algorithm using an optimization problem to minimize energy consumption.

Raca et al. in \cite{MagzineThroughputPrediction} presents the effects of the highly dynamic wireless communication on different application. Here they provide evidence of PHY layer metrics are used in assigning resources to the users in the cellular network. ABR is used as an example application to explain the advantages of using AI based techniques to learn accurate throughput prediction using PHY layer level metrics in cellular networks.

We classify the different ABR for wireless schemes into five main categories see Figure \ref{typesABR}. These categories are based on techniques used to design ABR or different types of objectives used in the design of the algorithms. Similarly the reviewed major approaches are summarized along with the references
in Table \ref{tab:summary}.

\section{Open challenges and opportunities}
\label{sec:openChal}
\subsection{Improving video for developing world clients}
Mobile phone adoption is even more explosive in developing countries. It has reached more than 98\% recently \cite{ITU2019}. But most of these devices are low end devices with slow network like 2G \cite{IhsanIMC16}. But video is the most dominant type of traffic in these parts of the world. This creates an even higher level challenge for Internet video providers \cite{UsamaDevelop}.

\subsection{Providing effective ABR for developing world}
New services like Web Light and  Facebook’s Free Basics service are introduced to improve the Internet quality and availability. Now, Free basics service is expanded to over 60 countries \cite{FreeBas} across select cellular service providers. But these services do not handle video elegantly. Both of these services replacing videos with an image \cite{Ihsanwww20}.

In future, demand for better quality Internet video will increase more for these low resource developing world clients. It is a challenge to provide even bare minimum service to these clients \cite{ITU2019}. But one can use techniques like  Oboe in \cite{4}. There are some proposed schemes which incorporate device level characteristics to improve the selection of bitrate \cite{JumaniZafarVideo}. 

In the developing world, vast amount of data access and high speed both are a luxury. Most of the communication is through text-based media like posters and flyers. Most of the  users in the developing regions are illiterate and  resource-constraint in terms of poor connectivity and little exposure to the technology. 
Access to computer and laptops is also very limited. Most of the Internet access is through low-end cellphone with a low-end camera. There has been many novel applications designed to solve different developing world specific problems. Many of these applications depends on video based solutions \cite{CHI09, CHI15}. Video streaming patterns of community health workers in Africa is studied in  \cite{CHIVideo17}.They demonstrated the effectiveness of health videos and also presented lessons for projects seeking to use multimedia content in rural setting.

AudioCanvas \cite{lowcostvid} is an application created for rural developing regions. It can be used by  telephone as an audio information system. This system enable rural users to interact directly with their pictures and receive narration or description. 

In a recent paper \cite{JumaniZafarVideo} authors have studied  effects of memory pressure on video streaming applications. Their experiments suggest QoE of video streaming is significantly affected by the selecting higher nitrates in a low memory cellphone. On the other hand \cite{IhsanIMC16} presents a comprehensive measurement study of cell phones used by users in developing world. Dataset  used in \cite{IhsanIMC16} has less than 1\% of the cellphone users in the developing world have more than 2GB memory in there devices. This creates a challenge of designing  specialized ABRs for developing world.

In their measurements  Nexus 5 phone with 2GB memory led to frequent video player crashes when it plays a 1080p video.  To achieve high QoE under low-memory scenarios they proposed an new scheme DAVS which adapt the playback buffer size based on conditions based on device memory pressure. 

\subsection{Optimize the spectral efficiency for video traffic }
The consistent exponential growth of video traffic will increase in the future with the advent of 5G based IoT devices. WiFi alliance has recently approved WiFi-6 \cite{wifi6} (802.11ax) with the focus of high density WLANs. These new changes will lead to even more congestion in the available spectrum specially in the unlicensed domain. There have been many recent studies to understand and optimize wireless spectrum sharing between different technologies like LTE or 5G based cellular networks and WiFi in unlicensed bands \cite{11, 12, 13,14}. Some of them used a machine learning-based approach to optimize spectrum sharing \cite{15}. But most of these papers are optimizing at the network level. This leads to many lost opportunities specifically related to video streaming \cite{29}. 

\subsection{Improving the QoE in the presence of network handovers }
\begin{figure*}
%	\vspace{-0.2in}
	\centering
	\includegraphics[width=\textwidth,height=6cm]{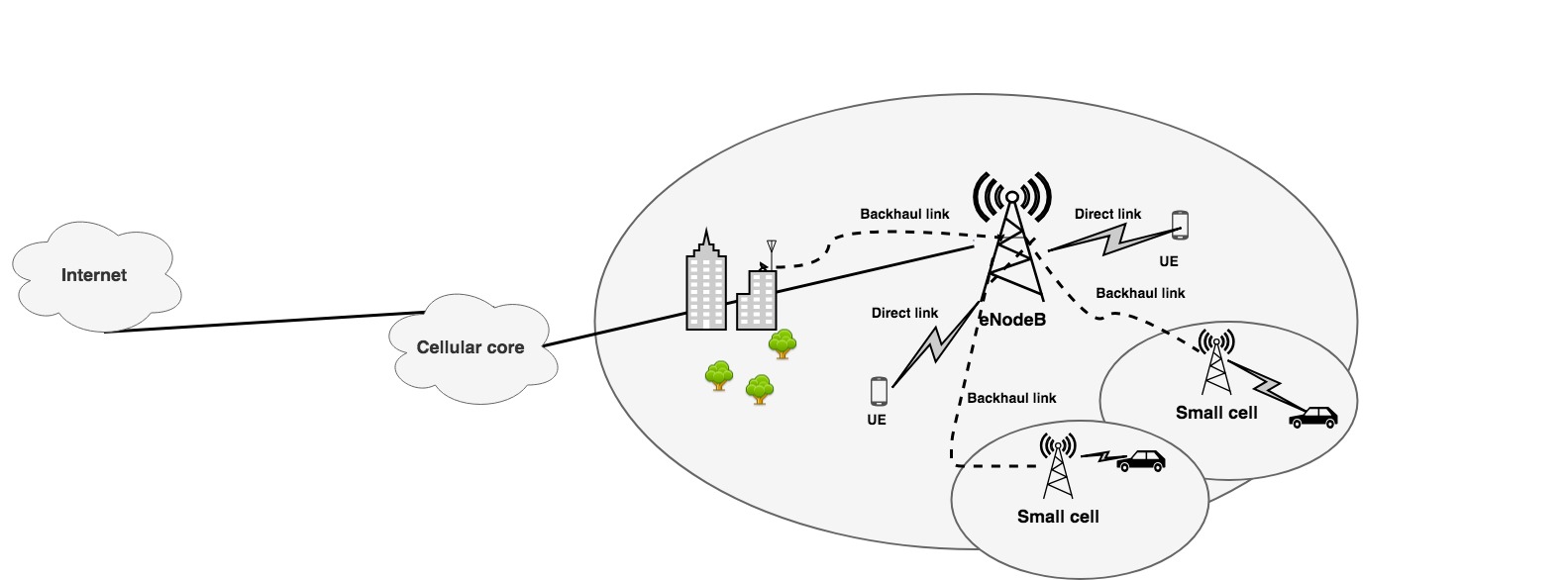}
	\vspace{0.2in}
		\centering
	\caption{Presence of Small Cells and frequent handovers in 5G}
	\label{fig:smallcell}
	%\vspace{-0.1in}
\end{figure*}

Some recent measurement studies \cite{24} shows that current policies of cellular carriers are not optimized, especially during handovers. These policies do not consider cell load information during handover resulting in degradation of application performance. In a 5G small cell, handovers will be more frequent. Figure \ref{fig:smallcell} shows a typical scenario of small cell based network in 5G vehicles in these small cells will experience frequent hand-offs. It will be critical for video streaming applications to perform well during these handovers.

\section{Open-source implementations of ABR algorithms and datasets}
\label{sec:opensource}
There are many open-source video players available on the Internet. But dash.js and ExoPlayer \cite{exoplayer} are the most popular in the industry and research.  ExoPlayer is developed by Google as the first  Android-based mobile DASH player. Many research paper have used it as their reference to implement their ABR algorithm. 
The other popular implementation is dash.js \cite{DASHIF}. It is developed by DASH Industry Forum which is supported by  most of the major players in Internet video industry like Akamai \cite{Akamai-dash}.
Previously GPAC \cite{gpac} was also very popular in research for prototyping. It is now called MP4Client. Both GPAC and DASH are implemented in JavaScript. 
In comparison to all the open-source implementations, dash.js encapsulates the standard and best practices. It is easy to customize and there is an Akamai reference implementation also available online which makes it easy to test. There are libraries available to use this for trace-driven analysis. Video providers wishing to use DASH often use the reference client dash.js to build their own video players. Table \ref{tbl:opensource} shows open-source implementation used in different papers. It is evident that most popular implementation in research is dash.js. Even in wireless network based evaluations \cite{37, QUAD} it is more commonly used due to its flexibility and acceptance.
There are hundreds of different ABR algorithms implemented using these open-source players. One of the popular one is of Pensive and their data traces \cite{pensive-code}.
Recently NAS \cite{NAS} implementation is also available open-source \cite{NAS-code}. 

\begin{table}
\centering
\caption{Summary of open-source ABR implementations} 
\begin{tabular}{|c|L|}
\hline
Implementation  & Corresponding papers  \\
\hline
Dash player & {Pensive  \cite{3},Oboe  \cite{4},HotDASH \cite{HotDASH}, Bursttracker \cite{37}, QUAD \cite{QUAD} and NAS \cite{NAS} }\\ 
\hline
GPAC video player & {\cite{MP-DASH} and Pstream \cite{PiStream15}} \\ 
\hline 
ExoPlayer & {IncorpPred \cite{IncorpPred} and QUAD \cite{QUAD}} \\ 
\hline
Trace driven & {QARC  \cite{QARC}, Comyco \cite{10}, and ACAA\cite{19ContentAwareQoE}} \\
\hline
\end{tabular}
\label{tbl:opensource}
\end{table}

There are two open-source data sets available \cite{DatasetsAVC, Dataset2} for trace-based analysis and to train machine learning algorithms. They are available in multiple encoding. These encoding rates are comparative to the large CDN provider like Hulu, YouTube and Netflix.

The testbed used to collect measurements in \cite{DatasetsAVC} is based on “MP4Client”, a multimedia player based on GPAC \cite{gpac}. They utilised very well known animated videos like Elephant Dreams, Big Buck Bunny  etc. Their files are obtained as  920x1080 YUV files. 

Similarly the second dataset \cite{DatasetsAVC} is a trace of 4G dataset which is composed of key performance indicators (KPIs) from two major Irish cellular providers. They are collected with different mobility patterns like static, car, bus and train. It has a very diverse range of throughput from 0 to 173 Mbit/sec. These traces are generated from a well-known Android network monitoring applications. There are few limitations of this dataset first all of its samples are of 1sec duration. Also, there is no GPS information in this dataset. To supplement the limitations of the real dataset there is repository of synthetic dataset.
\section{Conclusion}
\label{sec:conc}
In this paper, we surveyed different key techniques in the area of Internet video for wireless networks. It was observed that many approaches used cross layer communications on the clients to improve the re-buffering, quality switching
and encoding related impairments of the mobile video. It is important to note that with the upcoming deployments of WiFi-6 and 5G new challenges will arise which require us to rethink the implementation of ABR algorithms to address these challenges. 

\begin{small}
\bibliography{paper}

\begin{thebibliography}{100}
\providecommand{\url}[1]{#1}
\csname url@samestyle\endcsname
\providecommand{\newblock}{\relax}
\providecommand{\bibinfo}[2]{#2}
\providecommand{\BIBentrySTDinterwordspacing}{\spaceskip=0pt\relax}
\providecommand{\BIBentryALTinterwordstretchfactor}{4}
\providecommand{\BIBentryALTinterwordspacing}{\spaceskip=\fontdimen2\font plus
\BIBentryALTinterwordstretchfactor\fontdimen3\font minus
  \fontdimen4\font\relax}
\providecommand{\BIBforeignlanguage}[2]{{%
\expandafter\ifx\csname l@#1\endcsname\relax
\typeout{** WARNING: IEEEtran.bst: No hyphenation pattern has been}%
\typeout{** loaded for the language `#1'. Using the pattern for}%
\typeout{** the default language instead.}%
\else
\language=\csname l@#1\endcsname
\fi
#2}}
\providecommand{\BIBdecl}{\relax}
\BIBdecl

\bibitem{1}
Cisco VNI: Global Mobile Data Traffic Forecast Update, 2016- 2021.

\bibitem{2}
\BIBentryALTinterwordspacing
CSandvine: The global internet phenomena report. [Online]. Available:
  \url{https://www.sandvine.com/press-releases/sandvine-releases-2019-global-internet-phenomena-report}
\BIBentrySTDinterwordspacing

\bibitem{6}
Y.~Huang, ``Confused, timid, and unstable: Picking a video streaming rate is
  hard.'' in \emph{ACM IMC, 2012}, 2012.

\bibitem{7}
V.~S. Florin~Dobrian, ``Understanding the impact of video quality on user
  engagement.'' in \emph{SIGCOMM, 2011.}, 2011.

\bibitem{3}
R.~N. Hongzi~Mao and M.~Alizadeh, ``Neural adaptive video streaming with
  pensieve,'' in \emph{ACM SIGCOMM 2017.}, 2017.

\bibitem{4}
Z.~Akhtar, Y.~S. Nam, R.~Govindan, S.~Rao, J.~Chen, E.~Katz-Bassett,
  B.~Ribeiro, J.~Zhan, and H.~Zhang., ``Oboe: auto-tuning video abr algorithms
  to network conditions.'' in \emph{ACM SIGCOMM 2018}, 2018.

\bibitem{8}
V.~Mnih, A.~P. Badia, M.~Mirza, A.~Graves, T.~Lillicrap, T.~Harley, D.~Silver,
  and K.~Kavukcuoglu, ``Asynchronous methods for deep reinforcement learning,''
  in \emph{Proceedings of The 33rd International Conference on Machine
  Learning}, ser. Proceedings of Machine Learning Research, M.~F. Balcan and
  K.~Q. Weinberger, Eds., vol.~48.\hskip 1em plus 0.5em minus 0.4em\relax New
  York, New York, USA: PMLR, 20--22 Jun 2016, pp. 1928--1937.

\bibitem{9}
V.~Mnih, D.~S. Koray~Kavukcuoglu, A.~A. Rusu, J.~Veness, M.~G. Bellemare,
  A.~Graves, M.~Riedmiller, A.~K. Fidjeland, and G.~Ostrovski, ``Human-level
  control through deep reinforcement learning.'' in \emph{Nature, 2015.}, 2016,
  pp. 1928--1937.

\bibitem{13}
S.~Mustafa, K.~A. Alam, B.~Khan, M.~H. Ullah, and P.~Touseef, ``Fair
  coexistence of lte and wifi-802.11 in unlicensed spectrum, a systematic
  literature review,'' in \emph{ACM ICFNDS 2019.}, 2019.

\bibitem{15}
M.~{Alsenwi}, I.~{Yaqoob}, S.~R. {Pandey}, Y.~K. {Tun}, A.~K. {Bairagi},
  L.~{Kim}, and C.~S. {Hong}, ``Towards coexistence of cellular and wifi
  networks in unlicensed spectrum: A neural networks based approach.'' in
  \emph{IEEE Access 2019}, 2019.

\bibitem{19}
M.~A. et~al., ``Towards coexistence of cellular and wifi networks in unlicensed
  spectrum: A neural networks based approach,'' in \emph{IEEE Access 2019},
  2019.

\bibitem{PiStream15}
X.~Xie, X.~Zhang, S.~Kumar, and L.~E. Li, ``Pistream: Physical layer informed
  adaptive video streaming over lte,'' in \emph{ACM MobiCom 2015}, 2015.

\bibitem{37}
A.~Balasingam, M.~Bansal, R.~Misra, K.~Nagaraj, R.~Tandra, S.~Katti, and
  A.~Schulman, ``Detecting if lte is the bottleneck with bursttracker,'' in
  \emph{ACM Mobicom 2019}, 2019.

\bibitem{276}
X.~Ran, H.~Chen, X.~Zhu, Z.~Liu, and Jiasi, ``Chen. deepdecision: A mobile deep
  learning framework for edge video analytics.'' in \emph{In Proc. IEEE
  International Conference on Computer Communications, 2018}, 2018.

\bibitem{566}
Y.~Yuan, X.~Liang, X.~Wang, D.-Y. Yeung, and A.~Gupta., ``Temporal dynamic
  graph lstm for action-driven video object detection.'' in \emph{In Proc. IEEE
  International Conference on Computer Vision (ICCV) 2017}, 2017.

\bibitem{283}
K.~C. Zongqing~Lu, Noor~Felemban and T.~L. Porta., ``Demo abstract: On-demand
  information retrieval from videos using deep learning in wireless networks.''
  in \emph{In Proc. IEEE/ACM Second International Conference on
  Internet-of-Things Design and Implementation (IoTDI)}, 2019.

\bibitem{291}
Q.~Zhang, M.~Lin, L.~T. Yang, Z.~Chen, and P.~Li., ``Energy-efficient
  scheduling for real-time systems based on deep q-learning model.'' in
  \emph{IEEE Transactions on Sustainable Computing, 2017.}, 2017.

\bibitem{519}
S.~Rallapalli, H.~Qiu, A.~Bency, S.~Karthikeyan, R.~Govindan, B.~Manjunath, and
  R.~Urgaonkar., ``Are very deep neural networks feasible on mobile devices.''
  in \emph{IEEE Transactions on Circuits and Systems for Video Technology,
  2016}, 2012.

\bibitem{Survey-1}
S.~E. M.~Seufert, M.~Slanina, T.~Zinner, T.~Hoßfeld, and P.~Tran-Gia, ``A
  survey on quality of experience of http adaptive streaming,'' in \emph{IEEE
  Commun. Surveys Tuts., 2015.}, 2015.

\bibitem{SurveyABRoverHttp}
A.~{Bentaleb}, B.~{Taani}, A.~C. {Begen}, C.~{Timmerer}, and R.~{Zimmermann},
  ``A survey on bitrate adaptation schemes for streaming media over http,'' in
  \emph{IEEE Communications Surveys Tutorials 2019}, 2019.

\bibitem{Survey-2}
V.~T. P.~Juluri and D.~Medhi, ``Measurement of quality of experience of
  video-on-demand services: A survey,'' in \emph{IEEE Commun. Surveys Tuts.,
  2016.}, 2016.

\bibitem{19AccessSurvey}
N.~{Barman} and M.~G. {Martini}, ``Qoe modeling for http adaptive video
  streaming–a survey and open challenges,'' in \emph{IEEE Access, 2019},
  2019.

\bibitem{surveyABR17}
J.~{Kua}, G.~{Armitage}, and P.~{Branch}, ``A survey of rate adaptation
  techniques for dynamic adaptive streaming over http,'' in \emph{IEEE
  Communications Surveys Tutorials, 2017}, vol.~19, no.~3, 2017, pp.
  1842--1866.

\bibitem{DLWirelessSurvey}
Q.~{Mao}, F.~{Hu}, and Q.~{Hao}, ``Deep learning for intelligent wireless
  networks: A comprehensive survey,'' in \emph{IEEE Communications Surveys
  Tutorials}, 2019.

\bibitem{surveyVideoQua15}
M.~Seufert, ``A survey on quality of experience of http adaptive streaming,''
  in \emph{IEEE Communications Surveys Tutorials 2015}, 2015.

\bibitem{surveyABR20-2}
A.~A. {Barakabitze}, N.~{Barman}, A.~{Ahmad}, S.~{Zadtootaghaj}, L.~{Sun},
  M.~G. {Martini}, and L.~{Atzori}, ``Qoe management of multimedia streaming
  services in future networks: A tutorial and survey,'' in \emph{IEEE
  Communications Surveys Tutorials, 2020}, vol.~22, no.~1, 2020, pp. 526--565.

\bibitem{surveyABR19}
A.~{Bentaleb}, B.~{Taani}, A.~C. {Begen}, C.~{Timmerer}, and R.~{Zimmermann},
  ``A survey on bitrate adaptation schemes for streaming media over http,'' in
  \emph{IEEE Communications Surveys Tutorials, 2019}, 2019.

\bibitem{11}
L.~Bertizzolo, E.~Demirors, Z.~Guan, and T.~Melodia, ``Cobeam:
  Beamforming-based spectrum sharing with zero cross-technology signaling for
  5g wireless networks,'' in \emph{IEEE Infocom 2020}, 2020.

\bibitem{12}
S.~Zheng, T.~Han, Y.~Jiang, and X.~Ge, ``Smart contract-based secure spectrum
  sharing in multi-operators wireless communication networks,'' in \emph{ArXiv
  2020}, 2020.

\bibitem{DASH11}
I.~Sodagar, ``The mpeg-dash standard for multimedia streaming over the
  internet,'' in \emph{IEEE Multimedia, 2011.}, 2011.

\bibitem{AppleLive}
\BIBentryALTinterwordspacing
``Apple, “http live streaming.” [online]. available:.'' [Online].
  Available: \url{https://developer.apple.com/streaming}
\BIBentrySTDinterwordspacing

\bibitem{MicroSoftVideo}
\BIBentryALTinterwordspacing
``Microsoft, “microsoft smooth streaming.'' [Online]. Available:
  \url{http://www.iis.net/downloads/microsoft/smoothstreaming.}
\BIBentrySTDinterwordspacing

\bibitem{AdobeLive}
\BIBentryALTinterwordspacing
``Adobe, “http dynamic streaming,”.'' [Online]. Available:
  \url{http://www.adobe.com/products/hds-dynamicstreaming.html.}
\BIBentrySTDinterwordspacing

\bibitem{VeyasABR}
V.~S. Xiaoqi~Yin, Abhishek~Jindal and B.~Sinopoli., ``A control theoretic
  approach for dynamic adaptive video streaming over http.'' in \emph{In
  Proceedings of the ACM Conference on Special Interest Group on Data
  Communication, SIGCOMM, 2015.}, 2015.

\bibitem{BOLA}
R.~U. Kevin~Spiteri and R.~K. Sitaraman., ``Bola: Near-optimal bitrate
  adaptation for online videos.'' in \emph{In Proceedings of the IEEE
  International Conference on Computer Communications, INFOCOM, 2016}, 2016.

\bibitem{dash}
\BIBentryALTinterwordspacing
DASH Industry Forum. [Online]. Available:
  \url{https://github.com/Dash-Industry-Forum/dash.js}
\BIBentrySTDinterwordspacing

\bibitem{DashMMSys16}
L.~T. Federico~Chiariotti, Stefano~D’Aronco and P.~Frossard., ``Online
  learning adaptation strategy for dash clients.'' in \emph{In Proceedings of
  the International Conference on Multimedia Systems, MMSys 2016.}, 2016.

\bibitem{learningABR2}
M.~Claeys, S.~Latré, J.~Famaey, T.~Wu, W.~V. Leekwijck, and F.~D. Turck.,
  ``Design and optimisation of a (fa)q-learning-based http adaptive streaming
  client.'' in \emph{Connection Science, 26(1):25–43, 2014.}, 2014.

\bibitem{learningABR3}
J.~C. Virginia~Martín and N.~García., ``Design, optimization and evaluation
  of a q-learning http adaptive streaming client.'' in \emph{IEEE Transactions
  on Consumer Electronics, 62(4):380–388, 2016}, 2016.

\bibitem{18}
C.~Zhang, P.~Patras, and H.~Haddadi, ``Deep learning in mobile and wireless
  networking a survey,'' in \emph{IEEE COMMUNICATIONS SURVEYS and TUTORIALS
  2019}, 2019.

\bibitem{RLSurvey}
K.~Arulkumaran, M.~P. Deisenroth, M.~Brundage, and A.~A. Bharath, ``A brief
  survey of deep reinforcement learning,'' in \emph{IEEE SIGNAL PROCESSING
  MAGAZINE, SPECIAL ISSUE ON DEEP LEARNING FOR IMAGE UNDERSTANDING}, 2017.

\bibitem{Zero-ratedQoE}
A.~{Ahmad} and L.~{Atzori}, ``Mno-ott collaborative video streaming in 5g: The
  zero-rated qoe approach for quality and resource management,'' in \emph{IEEE
  Transactions on Network and Service Management 2020}, vol.~17, no.~1, 2020,
  pp. 361--374.

\bibitem{10}
T.~Huang, C.~Zhou, R.-X. Zhang, C.~Wu, X.~Yao, and L.~Sun, ``Comyco:
  Quality-aware adaptive video streaming via imitation learning,'' in \emph{ACM
  MM 2019}, 2019.

\bibitem{MP-DASH}
\BIBentryALTinterwordspacing
B.~Han, F.~Qian, L.~Ji, and V.~Gopalakrishnan, ``Mp-dash: Adaptive video
  streaming over preference-aware multipath,'' in \emph{Proceedings of the 12th
  International on Conference on Emerging Networking EXperiments and
  Technologies}, ser. CoNEXT ’16.\hskip 1em plus 0.5em minus 0.4em\relax New
  York, NY, USA: Association for Computing Machinery, 2016, p. 129–143.
  [Online]. Available: \url{https://doi.org/10.1145/2999572.2999606}
\BIBentrySTDinterwordspacing

\bibitem{42}
Z.~Song, L.~Shangguan, and K.~Jamieson, ``Wi-fi goes to town: Rapid picocell
  switching for wireless transit networks.'' in \emph{SIGCOMM 2017}, 2017.

\bibitem{HotDASH}
S.~{Sengupta}, N.~{Ganguly}, S.~{Chakraborty}, and P.~{De}, ``Hotdash: Hotspot
  aware adaptive video streaming using deep reinforcement learning,'' in
  \emph{2018 IEEE 26th International Conference on Network Protocols (ICNP)},
  2018, pp. 165--175.

\bibitem{QARC}
H.~Tianchi, Z.~Rui-Xiao, Z.~Chao, and S.~Lifeng, ``Qarc: Video quality aware
  rate control for real-time video streaming based on deep reinforcement
  learning,'' in \emph{Proceedings of the 26th ACM International Conference on
  Multimedia}, ser. MM ’18, 2018.

\bibitem{qflow}
B.~Rajarshi, B.~Archana, R.~Desik, R.~Mason, S.~Srinivas, K.~Dileep, M.~R.~K.
  P., and D.~Amogh, ``Qflow: A reinforcement learning approach to high qoe
  video streaming over wireless networks,'' in \emph{Mobihoc ’19}, 2019.

\bibitem{NAS}
H.~Yeo, Y.~Jung, J.~Kim, J.~Shin, and D.~Han, ``Neural adaptive content-aware
  internet video delivery,'' in \emph{13th {USENIX} Symposium on Operating
  Systems Design and Implementation ({OSDI} 18)}, 2018.

\bibitem{IncorpPred}
\BIBentryALTinterwordspacing
D.~Raca, A.~H. Zahran, C.~J. Sreenan, R.~K. Sinha, E.~Halepovic, R.~Jana,
  V.~Gopalakrishnan, B.~Bathula, and M.~Varvello, ``Incorporating prediction
  into adaptive streaming algorithms: A qoe perspective,'' in \emph{Proceedings
  of the 28th ACM SIGMM Workshop on Network and Operating Systems Support for
  Digital Audio and Video}, ser. NOSSDAV ’18.\hskip 1em plus 0.5em minus
  0.4em\relax New York, NY, USA: Association for Computing Machinery, 2018, p.
  49–54. [Online]. Available: \url{https://doi.org/10.1145/3210445.3210457}
\BIBentrySTDinterwordspacing

\bibitem{Jigsaw}
G.~Baig, J.~He, M.~A. Qureshi, L.~Qiu, G.~Chen, P.~Chen, and Y.~Hu, ``Jigsaw:
  Robust live 4k video streaming,'' in \emph{The 25th Annual International
  Conference on Mobile Computing and Networking}, ser. MobiCom ’19, 2019.

\bibitem{TMC19Hardware-assisted}
J.~{Yoon} and S.~{Banerjee}, ``Hardware-assisted, low-cost video transcoding
  solution in wireless networks,'' in \emph{IEEE Transactions on Mobile
  Computing, 2020}, 2020.

\bibitem{CASTLEAir19}
J.~Lee, J.~Lee, Y.~Im, S.~Dhawaskar~Sathyanarayana, P.~Rahimzadeh, X.~Zhang,
  M.~Hollingsworth, C.~Joe-Wong, D.~Grunwald, and S.~Ha, ``Castle over the air:
  Distributed scheduling for cellular data transmissions,'' in
  \emph{Proceedings of the 17th Annual International Conference on Mobile
  Systems, Applications, and Services}, ser. MobiSys ’19, 2019.

\bibitem{19ContentAwareQoE}
S.~Hu, M.~Xu, H.~Zhang, C.~Xiao, and C.~Gui, ``Affective content-aware
  adaptation scheme on qoe optimization of adaptive streaming over http,'' in
  \emph{ACM Transactions on Multimedia Computing, Communications, and Appli.
  2019}, 2019.

\bibitem{LinkForecast}
C.~{Yue}, R.~{Jin}, K.~{Suh}, Y.~{Qin}, B.~{Wang}, and W.~{Wei},
  ``Linkforecast: Cellular link bandwidth prediction in lte networks,'' in
  \emph{IEEE Transactions on Mobile Computing}, 2018.

\bibitem{QUAD}
Y.~Qin, S.~Hao, K.~R. Pattipati, F.~Qian, S.~Sen, B.~Wang, and C.~Yue,
  ``Quality-aware strategies for optimizing abr video streaming qoe and
  reducing data usage,'' in \emph{Proceedings of the 10th ACM Multimedia
  Systems Conference}, ser. MMSys ’19, 2019.

\bibitem{PowerRLforVBR}
C.~{Ye}, M.~C. {Gursoy}, and S.~{Velipasalar}, ``Power control for wireless vbr
  video streaming: From optimization to reinforcement learning,'' in \emph{IEEE
  Transactions on Communications}, 2019.

\bibitem{PiTree}
Z.~Meng, J.~Chen, Y.~Guo, C.~Sun, H.~Hu, and M.~Xu, ``Pitree: Practical
  implementation of abr algorithms using decision trees,'' in \emph{Proceedings
  of the 27th ACM International Conference on Multimedia 2019}, ser. MM ’19,
  2019.

\bibitem{16}
J.~Molina, D.~Muelas, J.~E.~L. de~Vergara, and J.~J.~G. Aranda, ``Network
  quality-aware architecture for adaptive video streaming from drones,'' in
  \emph{IEEE Internet Computing. 2020}, 2020.

\bibitem{17}
R.~Bhattacharyya, A.~Bura, D.~Rengarajan, M.~Rumuly, S.~Shakkottai,
  D.~Kalathil, R.~K.~P. Mok, and A.~Dhamdhere., ``Qflow: A reinforcement
  learning approach to high qoe video streaming over wireless networks.'' in
  \emph{ACM Mobihoc 2019}, 2019.

\bibitem{EdgeBilal}
K.~{Bilal} and A.~{Erbad}, ``Edge computing for interactive media and video
  streaming,'' in \emph{2017 Second International Conference on Fog and Mobile
  Edge Computing (FMEC)}, 2017, pp. 68--73.

\bibitem{EdgeShareARAR19}
\BIBentryALTinterwordspacing
X.~Ran, C.~Slocum, M.~Gorlatova, and J.~Chen, ``Sharear:
  Communication-efficient multi-user mobile augmented reality,'' in
  \emph{Proceedings of the 18th ACM Workshop on Hot Topics in Networks}, ser.
  HotNets ’19.\hskip 1em plus 0.5em minus 0.4em\relax New York, NY, USA:
  Association for Computing Machinery, 2019, p. 109–116. [Online]. Available:
  \url{https://doi.org/10.1145/3365609.3365867}
\BIBentrySTDinterwordspacing

\bibitem{EdgeFlexStream}
\BIBentryALTinterwordspacing
I.~Ben~Mustafa, T.~Nadeem, and E.~Halepovic, ``Flexstream: Towards flexible
  adaptive video streaming on end devices using extreme sdn,'' in
  \emph{Proceedings of the 26th ACM International Conference on Multimedia,
  2018}, ser. MM ’18.\hskip 1em plus 0.5em minus 0.4em\relax New York, NY,
  USA: Association for Computing Machinery, 2018, p. 555–563. [Online].
  Available: \url{https://doi.org/10.1145/3240508.3240676}
\BIBentrySTDinterwordspacing

\bibitem{EdgeBlockChain}
M.~{Liu}, F.~R. {Yu}, Y.~{Teng}, V.~C.~M. {Leung}, and M.~{Song}, ``Distributed
  resource allocation in blockchain-based video streaming systems with mobile
  edge computing,'' in \emph{IEEE Transactions on Wireless Communications,
  2019}, vol.~18, no.~1, 2019, pp. 695--708.

\bibitem{EdgeVideoCaching}
T.~X. {Tran} and D.~{Pompili}, ``Adaptive bitrate video caching and processing
  in mobile-edge computing networks,'' in \emph{IEEE Transactions on Mobile
  Computing, 2019}, 2019.

\bibitem{EdgeExtremeSDN}
M.~{Uddin}, T.~{Nadeem}, and S.~{Nukavarapu}, ``Extreme sdn framework for iot
  and mobile applications flexible privacy at the edge,'' in \emph{2019 IEEE
  International Conference on Pervasive Computing and Communications (PerCom},
  2019.

\bibitem{DataPlanStat}
\BIBentryALTinterwordspacing
``Ericsson. 2017. ericsson mobility report. (2017).'' [Online]. Available:
  \url{https://goo.gl/mjkwSH.}
\BIBentrySTDinterwordspacing

\bibitem{QualityAwareTrafficOffloading}
W.~Hu and G.~Cao, ``Quality-aware traffic offloading in wireless networks,'' in
  \emph{Proceedings of the 15th ACM International Symposium on Mobile Ad Hoc
  Networking and Computing}, ser. MobiHoc ’14, 2014, p. 277–286.

\bibitem{traceDASH}
R.~S. Kevin~Spiteri and D.~Sparacio., ``From theory to practice: Improving
  bitrate adaptation in the dash reference player.'' in \emph{In Proceedings of
  the 9th ACM Multimedia Systems Conference (MMSys 2018).}, 2018.

\bibitem{active1}
\BIBentryALTinterwordspacing
T.~Mangla, E.~Zegura, M.~Ammar, E.~Halepovic, K.-W. Hwang, R.~Jana, and
  M.~Platania, ``Videonoc: Assessing video qoe for network operators using
  passive measurements,'' in \emph{Proceedings of the 9th ACM Multimedia
  Systems Conference}, ser. MMSys ’18.\hskip 1em plus 0.5em minus 0.4em\relax
  New York, NY, USA: Association for Computing Machinery, 2018, p. 101–112.
  [Online]. Available: \url{https://doi.org/10.1145/3204949.3204956}
\BIBentrySTDinterwordspacing

\bibitem{ConfusedIMC12}
T.-Y. Huang, N.~Handigol, B.~Heller, N.~McKeown, and R.~Johari, ``Confused,
  timid, and unstable: Picking a video streaming rate is hard,'' in
  \emph{Proceedings of the 2012 Internet Measurement Conference}, 2012.

\bibitem{passive1}
I.~{Orsolic}, D.~{Pevec}, M.~{Suznjevic}, and L.~{Skorin-Kapov}, ``Youtube qoe
  estimation based on the analysis of encrypted network traffic using machine
  learning,'' in \emph{2016 IEEE Globecom Workshops (GC Wkshps)}, 2016, pp.
  1--6.

\bibitem{21}
Z.~W. Zhengfang~Duane, Abdul~Rehman, ``A quality-of-experience database for
  adaptive video streaming,'' in \emph{IEEE Transactions on Broadcasting 2018},
  2018.

\bibitem{22}
F.~Y. Yan, H.~Ayers, C.~Zhu, and K.~Winstein, ``Learning in situ: a randomized
  experiment in video streaming,'' in \emph{USENIX NSDI 2020}, 2020.

\bibitem{ABRoverQUICMeas}
D.~Bhat, A.~Rizk, and M.~Zink, ``Not so quic: A performance study of dash over
  quic,'' in \emph{Proceedings of the 27th Workshop on Network and Operating
  Systems Support for Digital Audio and Video}, ser. NOSSDAV’17, 2017.

\bibitem{MobileMeasre1}
K.~{Abdullah}, N.~{Korany}, A.~{Khalafallah}, A.~{Saeed}, and A.~{Gaber},
  ``Characterizing the effects of rapid lte deployment: A data-driven
  analysis,'' in \emph{2019 Network Traffic Measurement and Analysis Conference
  (TMA)}, 2019, pp. 97--104.

\bibitem{CSIMeasureABR}
\BIBentryALTinterwordspacing
S.~Xu, S.~Sen, and Z.~M. Mao, ``Csi: Inferring mobile abr video adaptation
  behavior under https and quic,'' in \emph{Proceedings of the Fifteenth
  European Conference on Computer Systems}, ser. EuroSys ’20.\hskip 1em plus
  0.5em minus 0.4em\relax New York, NY, USA: Association for Computing
  Machinery, 2020. [Online]. Available:
  \url{https://doi.org/10.1145/3342195.3387558}
\BIBentrySTDinterwordspacing

\bibitem{Flare360Video}
F.~Qian, B.~Han, Q.~Xiao, and V.~Gopalakrishnan, ``Flare: Practical
  viewport-adaptive 360-degree video streaming for mobile devices,'' in
  \emph{Proceedings of the 24th Annual International Conference on Mobile
  Computing and Networking}, ser. MobiCom ’18, 2018.

\bibitem{Rubiks360Video}
J.~He, M.~A. Qureshi, L.~Qiu, J.~Li, F.~Li, and L.~Han, ``Rubiks: Practical
  360-degree streaming for smartphones,'' in \emph{Proceedings of the 16th
  Annual International Conference on Mobile Systems, Applications, and
  Services}, 2018.

\bibitem{h264}
\BIBentryALTinterwordspacing
``2017. h264. (2017).'' [Online]. Available:
  \url{https://www.itu.int/rec/T-REC-H.264}
\BIBentrySTDinterwordspacing

\bibitem{hevc}
\BIBentryALTinterwordspacing
``2017. hevc. (2017).'' [Online]. Available:
  \url{https://www.itu.int/rec/T-REC-H.265}
\BIBentrySTDinterwordspacing

\bibitem{DRL360}
Y.~{Zhang}, P.~{Zhao}, K.~{Bian}, Y.~{Liu}, L.~{Song}, and X.~{Li}, ``Drl360:
  360-degree video streaming with deep reinforcement learning,'' in \emph{IEEE
  INFOCOM 2019 - IEEE Conference on Computer Communications}, 2019.

\bibitem{OnlineBitrate360}
\BIBentryALTinterwordspacing
M.~Tang and V.~W. Wong, ``Online bitrate selection for viewport adaptive
  360-degree video streaming,'' in \emph{CoRR 2020}, 2020. [Online]. Available:
  \url{https://arxiv.org/pdf/2005.02479.pdf}
\BIBentrySTDinterwordspacing

\bibitem{360MultiUser}
\BIBentryALTinterwordspacing
C.~Perfecto, M.~S. ElBamby, J.~D. Ser, and M.~Bennis, ``Taming the latency in
  multi-user vr 360-degree: A qoe-aware deep learning-aided multicast
  framework,'' in \emph{CoRR}, 2020. [Online]. Available:
  \url{http://arxiv.org/abs/1811.07388}
\BIBentrySTDinterwordspacing

\bibitem{25}
S.~M.~A. {Kazmi}, T.~N. {Dang}, I.~{Yaqoob}, A.~{Ndikumana}, E.~{Ahmed},
  R.~{Hussain}, and C.~S. {Hong}, ``Infotainment enabled smart cars: A joint
  communication, caching, and computation approach,'' in \emph{IEEE
  Transactions on Vehicular Technology 2019.}, 2019.

\bibitem{26}
S.~S. H.~Khan and M.~Bennis, ``Enhancing video streaming in vehicular networks
  via resource slicing,'' in \emph{IEEE Transactions on Vehicular Technology.
  2019}, 2019.

\bibitem{27}
Z.~Zhang, Y.~Yang, Y.~H. M.~Hua, C.~Li, and L.~Yang, ``Proactive caching for
  vehicular multi-view 3d video streaming via deep reinforcement learning,'' in
  \emph{IEEE Transactions on Wireless Communications, 2019}, 2019.

\bibitem{29}
R.Ali, B.Kim, S.W.Kim, and H.S.Kim, ``(relbt): A reinforcement learning-enabled
  listen before talk mechanism for lte-laa and wi-fi coexistence in iot,'' in
  \emph{Computer Communications 2020.}, 2020.

\bibitem{30}
D.~R.~C. Li, B.~Lin, and M.~Pan, ``Energy-efficient proactive caching for
  adaptive video streaming via data-driven optimization,'' in \emph{IEEE
  Internet of Things Journal, 2020}, 2020.

\bibitem{31}
H.~Shamim, H.~Zhang, C.~Moussa, and K.~Liu., ``Effective capacity-aware
  resource allocation for 5g ultra-dense network with hybrid access mode.'' in
  \emph{RACS 2019}, 2019.

\bibitem{32}
C.~E. Thornton, R.~M. Buehrer, A.~F. Martone, and K.~D. Sherbondy,
  ``Experimental analysis of reinforcement learning techniques for spectrum
  sharing radar,'' in \emph{Arxiv 2020}, 2020.

\bibitem{28}
A.~S. Melissa~Licciardello, Maximilian~Grüner, ``Understanding video streaming
  algorithms in the wild,'' in \emph{PAM 2020}, 2020.

\bibitem{41}
X.~Wang, L.~Kong, J.~Wu, X.~Gao, H.~Wang, and G.~Chen., ``Mmhandover: a
  pre-connection based handover protocol for 5g millimeter wave vehicular
  networks.'' in \emph{ACM IWQoS 2019.}, 2019.

\bibitem{45}
H.~Ge, X.~Wen, Z.~L. W.~Zheng, and B.~Wang., ``A history-based hand-dover
  prediction for lte systems.'' in \emph{International Symposium on
  ComputerNetwork and Multimedia Technology 2009}, 2009.

\bibitem{43}
I.~M. Bălan, B.~Sas, T.~Jansen, K.~S. I.~Moerman, and P.~Demeester., ``An
  enhanced weighted performance-based handover parameter optimization algorithm
  for lte networks.'' in \emph{EURASIP Journal on Wireless Communications
  andNetworking 2011}, 2011.

\bibitem{44}
W.~Luo, M.~C. X.~Fang, and X.~Zhou., ``An optimized handover trigger scheme in
  lte systems for high-speed railway.'' in \emph{In Proceedings of the Fifth
  In-ternational Workshop on Signal Design and Its Applications in
  Communications 2018}, 2018.

\bibitem{MMsys19LTEMesurement}
D.~Raca, A.~H. Zahran, C.~J. Sreenan, R.~K. Sinha, E.~Halepovic, R.~Jana,
  V.~Gopalakrishnan, B.~Bathula, and M.~Varvello, ``Empowering video players in
  cellular: Throughput prediction from radio network measurements,'' in
  \emph{ACM MMSys ’19}, 2019.

\bibitem{MLTCPthr}
M.~{Mirza}, J.~{Sommers}, P.~{Barford}, and X.~{Zhu}, ``A machine learning
  approach to tcp throughput prediction,'' in \emph{IEEE/ACM Transactions on
  Networking 2010}, vol.~18, no.~4, 2010, pp. 1026--1039.

\bibitem{MMSys20}
D.~Raca, D.~Leahy, C.~S. J., and J.~J. Quinlan, ``Beyond throughput: the next
  generation a 5g dataset with channel and context metrics,'' in \emph{ACM
  Multimedia Systems Conference}, ser. MMSys’ ’20, 2020.

\bibitem{ns3}
N.~Baldo, M.~Miozzo, M.~Requena-Esteso, and J.~Nin-Guerrero, ``An open source
  product-oriented lte network simulator based on ns-3. in proceedings of the
  14th acm international conference on modeling, analysis and simulation of
  wireless and mobile systems,'' in \emph{MSWiM 2011}, 2011.

\bibitem{VideoCrystalBall}
\BIBentryALTinterwordspacing
T.~Mangla, N.~Theera-Ampornpunt, M.~Ammar, E.~Zegura, and S.~Bagchi, ``Video
  through a crystal ball: Effect of bandwidth prediction quality on adaptive
  streaming in mobile environments,'' in \emph{Proceedings of the 8th
  International Workshop on Mobile Video}, ser. MoVid ’16.\hskip 1em plus
  0.5em minus 0.4em\relax New York, NY, USA: Association for Computing
  Machinery, 2016. [Online]. Available:
  \url{https://doi.org/10.1145/2910018.2910653}
\BIBentrySTDinterwordspacing

\bibitem{HotMobileCrossLayer}
\BIBentryALTinterwordspacing
F.~Lu, H.~Du, A.~Jain, G.~M. Voelker, A.~C. Snoeren, and T.~andreas, ``Cqic:
  Revisiting cross-layer congestion control for cellular networks,'' in
  \emph{Proceedings of the 16th International Workshop on Mobile Computing
  Systems and Applications}, ser. HotMobile ’15.\hskip 1em plus 0.5em minus
  0.4em\relax New York, NY, USA: Association for Computing Machinery, 2015.
  [Online]. Available: \url{https://doi.org/10.1145/2699343.2699345}
\BIBentrySTDinterwordspacing

\bibitem{Dataset2}
\BIBentryALTinterwordspacing
D.~Raca, J.~J. Quinlan, A.~H. Zahran, and C.~J. Sreenan, ``Beyond throughput: A
  4g lte dataset with channel and context metrics,'' in \emph{Proceedings of
  the 9th ACM Multimedia Systems Conference}, ser. MMSys ’18.\hskip 1em plus
  0.5em minus 0.4em\relax New York, NY, USA: Association for Computing
  Machinery, 2018, p. 460–465. [Online]. Available:
  \url{https://doi.org/10.1145/3204949.3208123}
\BIBentrySTDinterwordspacing

\bibitem{PensieveSIGCOMM}
HongziMao, R.~Netravali, and M.~Alizadeh, ``Neural adaptive video streaming
  with pensieve,'' in \emph{SIGCOMM ’17}, 2017.

\bibitem{24}
M.~Z. Xu~S., Nikravesh~A., ``Leveraging context-triggered measurements to
  characterize lte handover performance.'' in \emph{PAM 2019.}, 2019.

\bibitem{MagzineThroughputPrediction}
D.~{Raca}, A.~H. {Zahran}, C.~J. {Sreenan}, R.~K. {Sinha}, E.~{Halepovic},
  R.~{Jana}, and V.~{Gopalakrishnan}, ``On leveraging machine and deep learning
  for throughput prediction in cellular networks: Design, performance, and
  challenges,'' in \emph{IEEE Communications Magazine, 2020}, 2020.

\bibitem{EnergyAwareABR}
X.~{Chen}, T.~{Tan}, and G.~{Cao}, ``Energy-aware and context-aware video
  streaming on smartphones,'' in \emph{2019 IEEE 39th International Conference
  on Distributed Computing Systems (ICDCS)}, 2019, pp. 861--870.

\bibitem{ITU2019}
\BIBentryALTinterwordspacing
2019. Measuring digital development: Facts and figures 2019. [Online].
  Available: \url{https://www.itu.
  int/en/ITU-D/Statistics/Pages/facts/default.aspx.}
\BIBentrySTDinterwordspacing

\bibitem{IhsanIMC16}
S.~Ahmad, A.~L. Haamid, Z.~A. Qazi, Z.~Zhou, T.~Benson, and I.~A. Qazi, ``A
  view from the other side: Understanding mobile phone characteristics in the
  developing world,'' in \emph{Proceedings of the 2016 Internet Measurement
  Conference}, ser. IMC ’16.\hskip 1em plus 0.5em minus 0.4em\relax
  Association for Computing Machinery, 2016.

\bibitem{UsamaDevelop}
T.~B. Usama~Naseer, ``Configtron: Tackling network diversity with heterogeneous
  configurations.'' in \emph{USENIX HotCloud 2017}, 2017.

\bibitem{FreeBas}
\BIBentryALTinterwordspacing
Free Basics Platform. [Online]. Available:
  \url{https://developers.facebook.com/docs/internet-org.}
\BIBentrySTDinterwordspacing

\bibitem{Ihsanwww20}
A.~Tahir, M.~T. Munir, S.~M. Malik, Z.~A. Qazi, and I.~A. Qazi,
  ``Deconstructing google’s web light service,'' in \emph{Proceedings of The
  Web Conference 2020}, 2020.

\bibitem{JumaniZafarVideo}
A.~A. Jumani, F.~Zafar, Z.~A. Qazi, and I.~A. Qazi, ``Device-aware adaptive
  video streaming,'' in \emph{Proceedings of the ACM SIGCOMM 2019 Conference
  Posters and Demos}, New York, NY, USA, 2019.

\bibitem{CHI09}
D.~M. Frohlich, D.~Rachovides, K.~Riga, R.~Bhat, M.~Frank, E.~A. Edirisinghe,
  D.~Wickramanayaka, M.~Jones, and W.~Harwood, ``Storybank: mobile digital
  storytelling in a development context,'' in \emph{Proceedings of the 27th
  International Conference on Human Factors in Computing Systems, {CHI} 2009,
  Boston, MA, USA, April 4-9, 2009}.\hskip 1em plus 0.5em minus 0.4em\relax
  {ACM}, 2009, pp. 1761--1770.

\bibitem{CHI15}
N.~Kumar and R.~J. Anderson, ``Mobile phones for maternal health in rural
  india.'' in \emph{In Proceedings of the 33rd Annual ACM Conference on Human
  Factors in Computing Systems.}, 2015.

\bibitem{CHIVideo17}
\BIBentryALTinterwordspacing
M.~Molapo, M.~Densmore, and B.~DeRenzi, ``Video consumption patterns for first
  time smartphone users: Community health workers in lesotho,'' in
  \emph{Proceedings of the 2017 CHI Conference on Human Factors in Computing
  Systems}, ser. CHI ’17.\hskip 1em plus 0.5em minus 0.4em\relax New York,
  NY, USA: Association for Computing Machinery, 2017, p. 6159–6170. [Online].
  Available: \url{https://doi.org/10.1145/3025453.3025616}
\BIBentrySTDinterwordspacing

\bibitem{lowcostvid}
J.~Pearson, S.~Robinson, and M.~Jones, ``Exploring low-cost, internet-free
  information access for resource-constrained communities,'' in \emph{ACM
  Trans. Comput.-Hum. Interact., 2016}.\hskip 1em plus 0.5em minus 0.4em\relax
  Association for Computing Machinery, 2016.

\bibitem{wifi6}
\BIBentryALTinterwordspacing
``Wi-fi alliance® introduces wi-fi 6.'' [Online]. Available:
  \url{https://www.wi-fi.org/news-events/newsroom/wi-fi-alliance-introduces-wi-fi-6}
\BIBentrySTDinterwordspacing

\bibitem{14}
S.~Li, Y.~Huang, C.~Li, B.~A. Jalaian, Y.~T. Hou, and W.~Lou., ``Coping
  uncertainty in coexistence via exploitation of interference threshold
  violation.'' in \emph{ACM Mobihoc 2019}, 2019.

\bibitem{exoplayer}
\BIBentryALTinterwordspacing
``Google, “exoplayer.” [online]. available:.'' [Online]. Available:
  \url{https://google.github.io/ExoPlayer/guide.html}
\BIBentrySTDinterwordspacing

\bibitem{DASHIF}
\BIBentryALTinterwordspacing
``Dashif-reference-player. [online]. available:.'' [Online]. Available:
  \url{http://dashif.org/reference/players/javascript/1.4.0/samples/dash-if-reference-player/}
\BIBentrySTDinterwordspacing

\bibitem{Akamai-dash}
\BIBentryALTinterwordspacing
``Akamia player testing. [online]. available:.'' [Online]. Available:
  \url{http://players.akamai.com/players/dashjs}
\BIBentrySTDinterwordspacing

\bibitem{gpac}
\BIBentryALTinterwordspacing
``Gpac 2019. gpac multimedia open source project, javascript version of
  gpac’smp4box tool.'' [Online]. Available:
  \url{https://gpac.github.io/mp4box.js}
\BIBentrySTDinterwordspacing

\bibitem{pensive-code}
\BIBentryALTinterwordspacing
``H. mao, r. netravali, and m alizadeh. 2017. pensieve. code repository.''
  [Online]. Available: \url{https://github.com/hongzimao/pensieve}
\BIBentrySTDinterwordspacing

\bibitem{NAS-code}
\BIBentryALTinterwordspacing
``Nas. code repository.'' [Online]. Available:
  \url{https://github.com/kaist-ina/NAS_public}
\BIBentrySTDinterwordspacing

\bibitem{DatasetsAVC}
J.~J. Quinlan, A.~H. Zahran, and C.~J. Sreenan, ``Datasets for avc (h.264) and
  hevc (h.265) evaluation of dynamic adaptive streaming over http (dash),'' in
  \emph{Proceedings of the 7th International Conference on Multimedia Systems},
  ser. MMSys ’16, 2016.

\end{thebibliography}
\bibliographystyle{IEEEtran}
\end{small}

\begin{IEEEbiography}[{\includegraphics[width=1in,height=1.25in,clip,keepaspectratio]{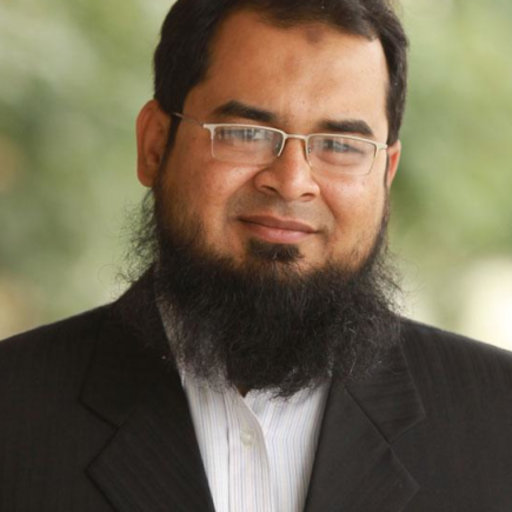}}]{Kamran Nishat} is a Postdoctoral Fellow in the Department of Engineering Technology, University of Houston, Texas, USA. He finished his PhD from  Department of Computer Science, LUMS, Lahore, Pakistan. He has received his B.Sc. degree in Mathematics in 1999 and MCS Computer Science degree, in 2001,from University of Karachi, Karachi. Prior to joining LUMS, he taught in the Department of Computer Science, University of Karachi, Karachi.He received his Masters in computer science in 2007 from LUMS. His research domain is Networking and Systems. He has published in leading Networking conferences including ACM CoNEXT and IEEE Infocom.
\end{IEEEbiography}

\begin{IEEEbiography}
    [{\includegraphics[width=1in,height=1.25in,clip,keepaspectratio]{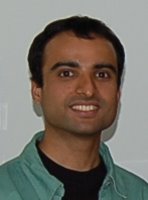}}]{Omprakash Gnawali} is an Associate Professor at the Computer Science Department of the University of Houston, USA. He does research in wireless networks, cybersecurity, and related technologies and has contributed research articles, open source software, and industry
standards. He received his SB and MEng from MIT, PhD from USC, and was a postdoc at Stanford. 
\end{IEEEbiography}

\begin{IEEEbiography}
    [{\includegraphics[width=1in,height=1.25in,clip,keepaspectratio]{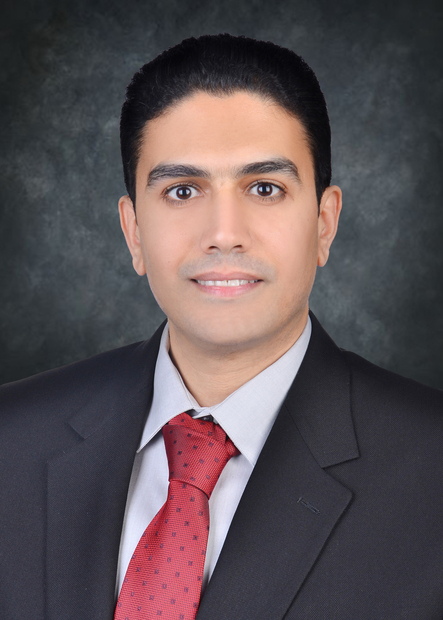}}]{Ahmed Abdelhadi} is an Assistant Professor at the University of Houston. Before joining UH, he was a Research Assistant Professor at Virginia Tech. He received his Ph.D. in Electrical and Computer Engineering from the University of Texas at Austin in 2011. He was a member in Wireless Networking and Communications Group (WNCG) and Laboratory of Informatics, Networks and Communications (LINC) group during his Ph.D. In 2012, he joined Bradley Department of Electrical and Computer Engineering and Hume Center for National Security and Technology at Virginia Tech. He was a faculty member of Wireless @ Virginia Tech. 
\end{IEEEbiography}

\end{document}